\documentclass[numbers]{cas-sc}
\usepackage[backend=bibtex,style=ieee,citestyle=numeric-comp,sorting=none,maxcitenames=2,mincitenames=1]{biblatex}
\usepackage{graphicx}
\usepackage{amsmath,amsthm, amssymb, float, bm, subcaption}
\usepackage{placeins}
\usepackage{longtable,tabularx,booktabs}
\usepackage{soul}
\usepackage{tcolorbox}
\tcbuselibrary{skins}
\usepackage{xcolor}
\usepackage[percent]{overpic}
\usepackage{miller}
\usepackage{tikz}
\usepackage{siunitx}
\newcommand{\fillin}[1]{\relax\ifmmode\mathbf{[#1]}\else\textbf{[#1]}\fi}
\definecolor{C0}{HTML}{1F77B4}
\definecolor{C1}{HTML}{FF7F0E}
\definecolor{C2}{HTML}{2ca02c}
\definecolor{C3}{HTML}{d62728}
\definecolor{C4}{HTML}{9467bd}
\definecolor{C5}{HTML}{8c564b}

\ifdefined\usetodonotes
\newlength{\todopane}
\usepackage[colorinlistoftodos,textsize=small]{todonotes}
\usepackage{marginnote}
\usepackage{zref-savepos,zref-abspage}
\makeatletter
\zref@addprop{savepos}{abspage}
\newcounter{reviewnote}
\newsavebox{\reviewnotebox}
\newlength{\reviewnoteshift}
\newlength{\reviewnotebottom}
\NewDocumentCommand{\reviewmarginpar}{o m}{%
  \stepcounter{reviewnote}%
  \edef\reviewnotelabel{reviewnote-\thereviewnote}%
  \zsavepos{\reviewnotelabel}%
  \sbox{\reviewnotebox}{\parbox[t]{\marginparwidth}{#2}}%
  \setlength{\reviewnoteshift}{\marginnotevadjust}%
  \edef\reviewnotepage{\zref@extractdefault{\reviewnotelabel}{abspage}{0}}%
  \ifnum\reviewnotepage>0\relax
    \ifcsname reviewbottom@\reviewnotepage\endcsname
      \setlength{\@tempdima}{\dimexpr
        \zposy{\reviewnotelabel}sp+\ht\reviewnotebox
        -\csname reviewbottom@\reviewnotepage\endcsname+6pt\relax}%
      \ifdim\@tempdima>\reviewnoteshift
        \setlength{\reviewnoteshift}{\@tempdima}%
      \fi
    \fi
    \setlength{\reviewnotebottom}{\dimexpr
      \zposy{\reviewnotelabel}sp-\reviewnoteshift-\dp\reviewnotebox\relax}%
    \expandafter\xdef\csname reviewbottom@\reviewnotepage\endcsname
      {\the\reviewnotebottom}%
  \fi
  \renewcommand*{\marginnotevadjust}{\reviewnoteshift}%
  \marginnote{\usebox{\reviewnotebox}}%
}
\NewDocumentCommand{\Todo}{O{} O{C2} O{0pt} m}{%
  \begingroup
  \renewcommand*{\marginnotevadjust}{#3}%
  \ifdefined\@captype
    \let\marginpar\reviewmarginpar
  \else
    \if@in@minipage@env
      \let\marginpar\reviewmarginpar
    \else
      \ifinner\let\marginpar\reviewmarginpar\fi
    \fi
  \fi
  \todo[color=#2!6,bordercolor=#2!6,linecolor=#2!45!black,
        size=\scriptsize,caption={#4}]{%
    \begin{tcolorbox}[
      width=\linewidth,boxsep=0pt,
      colback=#2!6,colframe=#2!45!black,
      colbacktitle=#2!45!black,coltitle=white,
      coltext=#2!25!black,
      fonttitle=\scriptsize\bfseries,fontupper=\scriptsize,
      title={#1},boxrule=0.4pt,arc=2pt,
      left=4pt,right=4pt,top=4pt,bottom=4pt,
      toptitle=3pt,bottomtitle=3pt,before skip=0pt,after skip=0pt
    ]%
      \raggedright #4
    \end{tcolorbox}%
  }%
  \endgroup
}
\makeatother
\NewDocumentCommand{\brandon}{O{} O{0pt} m}{%
  \IfBlankTF{#1}{\Todo[BR][C0][#2]{#3}}{\Todo[BR: #1][C0][#2]{#3}}%
}
\NewDocumentCommand{\abhijith}{O{} O{0pt} m}{%
  \IfBlankTF{#1}{\Todo[AA][C1][#2]{#3}}{\Todo[AA: #1][C1][#2]{#3}}%
}
\newcommand{\tododone}[1]{\todo[disable]{#1}}
\AddToHook{shipout/background}{%
  \begin{tikzpicture}[remember picture,overlay]
    \fill[black!7]
      ([xshift=-\todopane]current page.north east)
      rectangle (current page.south east);
    \draw[black!30,line width=0.6pt]
      ([xshift=-\todopane]current page.north east)
      -- ([xshift=-\todopane]current page.south east);
    \node[anchor=north,font=\scriptsize\bfseries\sffamily,text=black!45]
      at ([xshift=-0.75in,yshift=-0.18in]current page.north east)
      {TODO NOTES};
  \end{tikzpicture}%
}
\else
\NewDocumentCommand{\Todo}{O{} O{C2} O{0pt} m}{}%
\NewDocumentCommand{\brandon}{O{} O{0pt} m}{}%
\NewDocumentCommand{\abhijith}{O{} O{0pt} m}{}%
\newcommand{\tododone}[1]{}
\fi

\ifdefined\usechanges
\usetikzlibrary{calc}
\usepackage{fontawesome}
\ifdefined\usetodonotes\else
\fi
\usepackage[commentmarkup=todo,authormarkup=none]{changes} 
\definechangesauthor[color=C0]{R1} 
\definechangesauthor[color=C1]{R2} 
\definechangesauthor[color=C3]{R1,R2} 
\makeatletter
\@namedef{Changes@AuthorColor}{C3}
\colorlet{Changes@Color}{C3}
\renewcommand{\Changes@Markup@comment}[3]{%
  \IfStrEq{\Changes@optioncommentmarkup}{todo}%
		{\colorlet{Changes@todocolor}{authorcolor}\todo[color=Changes@todocolor!10, bordercolor=Changes@todocolor, linecolor=Changes@todocolor!70, nolist]{\faBookmarkO\ \textbf{#1}}}{}}
\makeatother
\newtcolorbox{addedbox}[3][]
{
  colframe=#2,
  colback=#2!10,
  boxrule=0pt,
  title={\ \ \ \ \faBookmarkO\ \textsf{\textbf{#3}}},
  boxsep=1mm,
  left=-1mm,
  right=-1mm,
  enhanced,
  width=\linewidth,
  #1,
}
\else
\NewDocumentEnvironment{addedbox}{ O{} m m }{}{}
\usepackage[final,nocomment]{changes}
\fi

\newtcolorbox{mybox}[1][]{colback=#1!10,colframe=#1!80,boxrule=1pt,arc=1mm,left=0.1mm,right=0.1mm,top=0.1mm,bottom=0.1mm}

\usepackage[capitalize]{cleveref}
\crefrangeformat{subfigure}{Fig.~#3#1#4--#5\crefstripprefix{#1}{#2}#6} 

\makeatletter
\define@key{Gin}{Crop3D}[]{\setkeys{Gin}{trim=21mm 15mm 80mm 22mm,clip}}
\makeatother

\makeatletter                                                                   
\newlength{\bibsep}{\@listi \global\bibsep\itemsep \global\advance\bibsep by\parsep} 
\makeatother 

\ExplSyntaxOn
\cs_set:Npn \__first_footerline:
{
  \group_begin:
  \small\sffamily
  \ifnum\theblind>0\relax
  \else
    \__short_authors:
  \fi
  \group_end:
}
\ExplSyntaxOff

\begin{document}

\shortauthors{Anantharanga, Runnels}  
\shorttitle{}
\title [mode = title]{A mesoscale damage model for void nucleation and growth in shock loaded microstructure informed by computer vision}
\author{Abhijith Thoopul Anantharanga}[orcid=0009-0009-6274-8771]
\author{Brandon Runnels}[orcid=0000-0003-3043-5227]
\cormark[1]
\cortext[1]{Corresponding author}
\ead{brunnels@iastate.edu}
\affiliation{organization={Department of Aerospace Engineering, Iowa State University},
city={Ames},
state={IA},
country={USA}}

\begin{abstract}
Spallation in ductile materials under high strain rates occurs through the process of void nucleation, growth, and coalescence.
Void nucleation in microstructures is known to depend on a wide range of features ranging from microstructural features to local stress state.
The criteria for void nucleation in damage models are highly stochastic, and microstructural awareness is often neglected in nucleation models. 
Recently, it was shown that void nucleation propensity can be captured using a computer vision technique.
In the present work, this is integrated with a continuum damage model framework for comprehensive simulation of spallation in tantalum under plate impact.
Void nucleation probability in this model is governed by a susceptibility field, predicted from grain orientations and grain boundary energy, combined with a mechanical barrier set by the local maximum principal stress. 
This is coupled to a dislocation density based crystal plasticity model within a Lagrangian shock simulation, in which the microstructure is selectively resolved to reduce computational cost (without alteration of bulk wave dynamics).
The model is calibrated at a flyer velocity of $\SI{196}{\meter\per\second}$, where the simulated free surface velocity (FSV) matches experimental measurements of shock arrival, peak state, release and the pullback signal measured in flyer impact experiments, with some deviations confined to the late time ringing.
With no further parameter changes, the model transfers to a $\SI{214}{\meter\per\second}$ case and accurately reproduces the FSV.
To demonstrate the microstructure-awareness of the nucleation model, a selection of synthetic microstructures is presented for comparison.
In the resolved microstructure, voids nucleate predominantly (though not exclusively) at grain boundaries and triple junctions.
It is shown that the FSV differs significantly based on the incorporated microstructure, yielding the correct FSV prediction only in the case of representative microstructures.
\end{abstract}

\maketitle

\section{Introduction}\label{sec:Introduction}
Structural materials in defense and aerospace applications are frequently subjected to high impact dynamic loads, arising during events such as armor penetration, ballistic impact, and hypervelocity collisions of debris with spacecraft \cite{remington2015microjoules}.
These loads subject the materials to high strain rates of the order of $\SIrange{e4}{e7}{\per\second}$ \cite{meyers1994dynamic,gray2012high}. 
Under these extreme conditions, ductile materials commonly fail by spallation resulting from void nucleation, growth, coalescence, and ultimately material fragmentation under a dynamic tensile load \cite{antoun2003spall,curran1987dynamic}.
The spall resistance of materials is known to depend on microstructural features such as misorientation, grain size and grain boundary properties such as grain boundary energy (GBE) and triple junction density \cite{yang2022grain,fensin2012influence,brown2015microstructural,escobedo2011effects}.
Designing microstructures that resist spallation therefore requires knowledge of where voids nucleate and how they grow and coalesce.

This effect is commonly studied experimentally using plate impact experiments \cite{antoun2003spall}.
The impact of the flyer with the target generates compressive shock waves in both the flyer and the target, which reflect from the free surfaces of both the flyer and the target as rarefaction fans \cite{antoun2003spall}.
When the flyer and target are made of the same material, the two rarefaction waves interact inside the target at a depth from its free surface approximately equal to the flyer thickness \cite{antoun2003spall}.
Their interaction gives rise to a high tensile stress in this region, and if this stress is high enough, void nucleation occurs. 
The region where voids nucleate and eventually grow and coalesce is called the spall plane.
Damage inside the target is difficult to observe during the experiment.
Optical diagnostics such as velocity interferometers are mostly limited to surface measurements \cite{luo2012gas}.
For \textit{in-situ} characterization,  X-ray techniques may be used to probe the interior of the sample. 
Single pulse X-ray phase contrast imaging at a synchrotron source has captured spallation inside a low-Z material impacted by a gas-gas driven plunger \cite{luo2012gas}. 
Ultra fast small angle X-ray scattering at an X-ray free electron laser has been able to quantify void nucleation, growth and coalescence in 1~\si{\micro\meter} thick copper films impacted by laser ablation \cite{coakley2020femtosecond}.
However, these techniques are extremely expensive to perform and have been demonstrated on low-Z materials and micrometer thick foils.
Damage progression is therefore mainly inferred from indirect measurements, often through the measurement of the velocity history of the free surface of the target recorded by laser interferometry \cite{jones2018effect,barker1972laser}.
The pullback signal, which is defined as the difference between the peak free surface velocity (FSV) and the subsequent velocity minimum, is related to the spall strength of the material \cite{jones2018effect}.
The oscillation in the FSV profile that follows the pullback carries implicit information about the average growth and coalescence of voids \cite{kanel2010spall}, though it is not sufficient to distinguish between different damage distributions within the spall plane \cite{jones2018effect}.
Post-mortem characterization of soft recovered targets helps resolve this ambiguity by showing the damage field in microstructures \cite{jones2018effect,escobedo2011effects}.
A useful and predictive spallation model should mimic this experimental protocol and be able to reproduce the FSV and also predict damage fields in the microstructure.

In addition to experiment, the spall process has been studied using a variety of computational methods at varying scales.
At the individual void level, voids can be resolved explicitly in molecular dynamics and void-resolved crystal plasticity simulations  \cite{seppala2004effect,tang2012ductile,lebensohn2013modeling}.
While these simulations capture void growth and interactions directly, their cost restricts them to domains far smaller than a plate impact experiment scale.
At the plate impact experiment scale, damage is often represented by continuum models that do not resolve individual voids.
Continuum damage models require a failure or nucleation criterion to predict void locations.
\textcite{tuler1968criterion} proposed a failure criterion in which fracture occurs once a threshold tension has been sustained for a critical time. 
This was followed by porosity based models in which void volume fraction evolves through separate nucleation and growth laws calibrated to experimental data  \cite{curran1987dynamic,johnson1981dynamic}.
Later models added the effects of micro inertia which becomes relevant at extreme strain rates \cite{molinari2005physical,czarnota2006modelling,wilkerson2014dynamic}.
Further, void growth models have been coupled with crystal plasticity, so that anisotropic plastic flow of the matrix surrounding the voids is captured \cite{nguyen2017dislocation}.
Phase field formulations have been applied to spallation with coupling to Gurson type void growth \cite{zhang2021phase,zhang2023phase,han2024study}.
While many models are able to capture the FSV \cite{czarnota2006modelling,schmelzer2025statistical}, models incorporating microstructure awareness (and specifically grain boundary energy) in void nucleation and growth remain limited.

The precise criterion for void nucleation is dependent on a number of factors ranging from microstructural features to the local stress state at the time of nucleation.
In most classical models, void nucleation is solely controlled by the local stress.
In the earliest criterion, fracture begins to occur once a time integral of the tension above a threshold value reaches a critical value \cite{tuler1968criterion}.
In porosity based models voids nucleate at a rate that increases with the tension above a threshold stress \cite{curran1987dynamic}, and some have void nucleation at nucleation pressures described by a prescribed probability distribution \cite{versino2018computationally}.
More recently, microstructure-aware nucleation models have been proposed.
Crystal plasticity simulations of tantalum polycrystals have been able to quantify stress heterogeneity at grain boundaries and combine it with nucleation strength at grain boundaries from molecular dynamics in a probabilistic nucleation criterion \cite{schmelzer2025statistical}.
Further, data driven reduced order models have been developed to predict the probability distribution of the local stress in polycrystals that leads to porosity based damage \cite{zhang2023data}.
A porosity based damage model has also been developed in which the nucleation resistance follows a probability distribution informed by molecular dynamics and polycrystal calculations \cite{schmelzer2025thermo}.
Deep learning models trained on crystal plasticity have been developed to predict spall response from microstructure, such as \cite{saha2026numerical}, though they do not account for grain boundary properties.
Given that voids appear to nucleate preferentially at certain grain boundaries \cite{fensin2014some}, it is essential to incorporate their properties into a nucleation model.

To address this issue of grain boundary aware nucleation, \textcite{anantharanga2026predicting} presented a model inspired by computer vision: a U-net type model, trained on grain orientation and GBE was shown to predict void nucleation susceptibility fields along grain boundaries in shock loaded body centered cubic (BCC) tantalum \cite{anantharanga2026predicting}.
The present work extends this computer vision model to account for the local stress state by coupling it to a full mesoscale plasticity model under shock loading.
The proposed technique combines Lagrangian shock physics, a dislocation density based crystal plasticity model, a continuum damage model for void growth, and the void nucleation criterion based on machine learned susceptibility fields predicted from grain boundary energy and grain orientations and local stress states.

The remainder of this paper is organized as follows.
\Cref{sec:shockhydrocode} presents the Lagrangian shock physics formulation, crystal plasticity, continuum damage, and nucleation models.
\Cref{sec:machine_learning} presents the machine learning framework used to predict void nucleation probability incorporated in this work.
\Cref{sec:domain,sec:implementation} describe the techniques used to develop the computational domain used in simulations and their implementation.
Finally, \cref{sec:Results} presents the FSV validation against plate impact experiments and shows void nucleation and damage growth in a grain resolved spall region, and \cref{sec:Conclusions} summarizes the findings and discusses directions for future work.

\section{Lagrangian shock physics}\label{sec:shockhydrocode}
This section presents the shock physics balance laws and models incorporated in this work.
The method is based on the standard Lagrangian formulation for shock propagation in solids \cite{wilkins1999}, which supplies the volumetric response of the material by an equation of state \cite{benson1992}.
Balance laws are first introduced, followed by a section on the equation of state and the artificial viscosity.
Subsequent sections introduce the crystal plasticity, damage, and nucleation models.
As with all models of this type, numerous parameters are required.
These are summarized for all model components, except for the computer vision model, in \cref{tab:params}.

The target and the flyer are both described in the Lagrangian frame, with each material point being labeled by its reference position $\mathbf{X}$.
The deformation gradient is  $\mathbf{F} = \partial \mathbf{x} / \partial \mathbf{X}$, where $\mathbf{x}$ is a point in the deformed configuration, with  $J = \det \mathbf{F}$.
The balance laws of mass, energy and momentum are taken in their standard form as used in most shock hydrocodes \cite{wilkins1999,benson1992}.
They are given by
\begin{equation}
    \rho J = \rho_0, \qquad \rho_0 \, \dot{\mathbf{v}} = \nabla_0\cdot \mathbf{P}, \qquad \rho_0 \, \dot{e} = \mathbf{P}:\dot{\mathbf{F}}\label{eq:conservation_eqn}
\end{equation}
where $\rho_0$ and $\rho$ are the reference density and current density, $\mathbf{v}$ is the velocity,  $\mathbf{P}$ is the first Piola-Kirchhoff stress, $e$ is the specific internal energy, $p$ is the pressure, and $:$ is the double contraction operation.
In the Lagrangian frame, the target-flyer interface, the free surfaces and the material microstructure therefore stay on the same grid points throughout the simulations and no interface tracking is needed.
Since each cell retains its reference mass, mass conservation is satisfied by construction.

The above differential form of conservation equations holds only when the fields are smooth, but across a shock, the particle velocity, density and internal energy jump  discontinuously.
Therefore the Rankine-Hugoniot jump conditions are required to model shock waves propagating in solid materials.
A brief statement of these jump conditions is given here but the reader is referred to the literature for complete derivations \cite{meyers1994dynamic, davison2008}.
For a steady shock that enters a material at rest with zero pressure the jump conditions are,
\begin{equation}\label{eq:rh}
    \rho_0 U_s = \rho(U_s-u_p),\qquad P_H =\rho_0 U_s u_p \qquad e_H - e_0 = \frac{1}{2}P_H(\frac{1}{\rho_0} - \frac{1}{\rho})
\end{equation}
where $U_s$ is the shock velocity, $u_p$ is the particle velocity behind the shock, $\rho_0$ and $\rho$ are the densities ahead of and behind the shock, $P_H$ is the pressure behind the shock, and $e_0$ and $e_H$ are the specific internal energies ahead of and behind the shock.
The relations \cref{eq:rh} express the conservation of mass, energy and momentum across the shock.

\subsection{Equation of state}\label{sec:equation_of_state}
The jump conditions are closed by an equation of state which relates the pressure to the density and internal energy.
For metals under shock loading, the Mie-Gr\"uneisen equation of state is often considered \cite{mie1903,clayton2019} and is used here:
\begin{equation}\label{eq:mg}
    p(\rho , e) = P_\mathrm{ref}(\rho) + \Gamma_0 \rho_0(e-e_\mathrm{ref}(\rho)\bigr),
\end{equation}
where $\Gamma_0$ is the Gr\"uneisen parameter, and the product $\Gamma \rho = \Gamma_0 \rho_0$ is a model constant for metals at moderate compression~\cite{meyers1994dynamic}.
The principal Hugoniot, which is the set of states reached by single shocks from the material at rest, is obtained from jump conditions \cref{eq:rh}.
It has a linear relation between shock velocity and particle velocity, $U_s = C_0 + s\,u_p$, where $C_0$, the bulk sound speed, and $s$, the slope of the relation between $U_s$ and $u_p$ are fitted to plate-impact data for tantalum~\cite{marsh1980}.
In terms of the volumetric strain,  $\eta = 1 - \rho_0/\rho$, the principal Hugoniot is,
\begin{equation}\label{eq:hugonoit}
    P_H(\eta) = \frac{\rho_0C_0^2 \eta}{(1 - s\eta)^2}, \qquad e_H(\eta) = \frac{P_H(\eta)\eta}{2\rho_0}.
\end{equation}

A single shock from the reference state lies on the Hugoniot with $e = e_H$, and \cref{eq:mg} with $P_\mathrm{ref} = P_H$ returns $p = P_H$ exactly.
States that lie off the Hugoniot such as those reached along release paths, carry an excess energy $e - e_H$, which the second term of \cref{eq:mg} converts into a thermal pressure.

However, the Hugoniot is defined only during compression.
A shock cannot be tensile, so for  $\eta < 0$ there is no Hugoniot to serve as the reference and there is no physical basis for \cref{eq:hugonoit}.
This is a limitation for the present work as the material at the spall plane is under tension. 
In the present approach, the reference curve is continued into tension as a linear elastic branch with the same zero-pressure bulk modulus,
\begin{equation}\label{eq:pref}
    P_\mathrm{ref}(\eta) = \begin{cases} P_H (\eta)  & \eta  \ge 0, \\ \rho_0 C_0^2\eta & \eta < 0 \end{cases}, \qquad e_\mathrm{ref}(\eta) = \frac{P_\mathrm{ref}(\eta) \eta}{2\rho_0}
\end{equation}
The compressive and tensile forms of \cref{eq:pref} vanish when $\eta = 0$ and they have the same slope $\rho_0 C_0^2$.
Therefore, the sound speed and the pressure remain continuous as the material passes from compression to tension.
The local sound speed follows from \cref{eq:mg} as
\begin{equation}\label{eq:soundspeed}
    c^2 =\left. \frac{\partial p}{\partial \rho}\right|_e + \frac{p}{\rho^2}\left. \frac{\partial p}{\partial e}\right|_\rho =\frac{\mathrm{d}P_{ref}}{\text{d}\rho} - \Gamma_0\rho_0\frac{\mathrm{d}e_{ref}}{\text{d}\rho} + \frac{\rho_0\Gamma_0 p}{\rho^2}
\end{equation}
(It is noted that the material parameters listed in \cref{tab:params} are separated from numerical parameters, since the former are calibrated against measured quantities.)

\subsection{Artificial viscosity}\label{sec:shock_av}
Any finite discretization of \cref{eq:conservation_eqn} produces spurious oscillations behind the shock front.
Further, the entropy that the shock must produce is not recovered accurately due to the presence of these oscillations \cite{vonneumann1950}.
The standard remedy to fix this issue in many Lagrangian hydrocodes is the use of an artificial viscosity that is added to the pressure \cite{vonneumann1950, landshoff1955}.
In one dimension, it takes the form,
\begin{equation}\label{eq:av1d}
    q = \begin{cases}  C_Q \rho (\Delta v)^2  + C_L \rho c |\Delta v|   &  \Delta v<0 , \\ 0 &  \Delta v \ge 0  \end{cases}
\end{equation}
where $\Delta v$ is the velocity jump across the cell, and $C_Q$ and $C_L$ are the quadratic and linear coefficients.
The quadratic term spreads the shock over a finite width that is set by $C_Q$ and is independent of the shock strength \cite{vonneumann1950}. 
The linear term damps the oscillations that the quadratic term alone leaves behind \cite{landshoff1955}.
It is important to note that since the scheme is conservative, the smeared shock connects states which satisfy \cref{eq:rh}, and the kinetic energy removed by $q$ reappears as internal energy \cite{wilkins1980, caramana1998}. 

However, \cref{eq:av1d} has two limitations in two dimensions.
First, it depends on a single velocity jump, which is not unique when the velocity varies in two directions \cite{wilkins1980, caramana1998}.
A common choice is to compute $q$ from $\nabla \cdot \mathbf{v}$ and apply it equally in all directions.
Under uniaxial compression, this adds a viscous pressure in the transverse direction, where there is no shock.
Second, \cref{eq:av1d} acts only in compression, therefore oscillations during tension are not damped.
Such oscillations arise near nucleated voids during release.
In the current approach, both limitations are addressed by the use of a directional viscous stress. 
Each normal component responds to the velocity jump in its own direction. 
\begin{equation} \label{eq:av2d}
    \mathbf{q} = \begin{bmatrix}q (\Delta v_x) & 0 \\0 & q (\Delta v_y)  \end{bmatrix} , \qquad   \Delta v_x = \frac{\partial v_x}{\partial x}\Delta x, \qquad \Delta v_y = \frac{\partial v_y}{\partial y}\Delta y 
\end{equation}
Here, $\Delta v_x$ and $\Delta v_y$ are the jumps in the normal velocity components.

\subsection{Crystal plasticity}\label{sec:crystalplasticity}
The equation of state computes the volumetric response of the material (\cref{sec:shockhydrocode}), but the deviatoric stress that drives grain boundary void nucleation depends on the elastic and plastic response of each grain in the polycrystal and this response is captured by a crystal plasticity model.
Crystal plasticity models commonly describe slip rates by a power law of the resolved shear stress, or by a thermally activated rate law with a fixed reference slip rate, with the slip resistance used as an internal variable \cite{asaro1983,kothari1998}.
However, in shock loaded materials the strain rate is limited by the density and velocity of mobile dislocations and the glide mechanism changes from thermal activation over obstacles to phonon drag \cite{austin2011}.
In the present work, a dislocation density based crystal plasticity model for BCC tantalum is incorporated \cite{nguyen2021dynamic,lee2026bayesian}.
The slip rate is related to the mobile dislocation density through the Orowan relation, and the glide velocity combines thermally activated waiting time at obstacles with drag-limited running time between them \cite{austin2011, luscher2017, nguyen2017dislocation}.
The model is summarized here but a detailed description is given in \cite{nguyen2021dynamic, lee2026bayesian}.

The usual multiplicative decomposition is used to describe the deformation:
given a time-dependent mapping $\bm{\varphi}:(\mathbb{R}^3,\mathbb{R})\to\mathbb{R}^3$, the deformation gradient is decomposed as
\begin{equation}\label{eq:Fdecomp}
    \mathbf{F} \; \stackrel{\text{def}}{=} \frac{\partial \varphi}{\partial \mathbf{X}} = \mathbf{F^e}\mathbf{F^p}
\end{equation}
where $\mathbf{F^p}$ is the (not necessarily compatible) plastic part of the deformation gradient in an intermediate configuration, and $\mathbf{F}^e$ is the elastic part of the deformation gradient. 
The spatial velocity gradient $\mathbf{L}$ is decomposed additively into the elastic distortion rate tensor $ \mathbf{L}^e$ and the plastic distortion rate $\mathbf{L}^p$ tensor as 
\begin{equation}\label{eq:Ldecomp}
\mathbf{L} \; \stackrel{\text{def}}{=} \frac{\partial v}{\partial \mathbf{x}} = \dot{\mathbf{F}}\mathbf{F^{-1}} = \mathbf{L^e} + \mathbf{F^e}\mathbf{L^p}\mathbf{F^{e-1}}, \qquad \mathbf{L^e} = \dot{ \mathbf{F^e}} \mathbf{F^{e-1}}, \qquad \mathbf{L^p} = \dot{ \mathbf{F^p}} \mathbf{F^{p-1}}
\end{equation}
where $v$ is the spatial velocity field. 
Plastic flow is the sum of the dislocation glide on $n_{\text{slip}}$ slip systems so that
\begin{equation}\label{eq:Lp}
\mathbf{L^p} = \sum_{\alpha=1}^{n_{slip}}\dot{\gamma}_p^{\alpha}\mathbf{S_0^\alpha}, \qquad \mathbf{S}_0^{\alpha}=\mathbf{s}_0^{\alpha} \otimes \mathbf{m}_0^{\alpha}  
\end{equation}
where $\dot{\gamma^\alpha}$ is the slip rate on system $\alpha$ and $\mathbf{S}_0^\alpha$ is its Schmid tensor$\alpha$, and $\mathbf{s}_0^\alpha$ and $\mathbf{m}_0^\alpha$ are the unit slip direction and slip plane normal in the intermediate configuration, respectively. 
The elastic Green-Lagrange strain measured in the intermediate configuration is
\begin{equation}\label{eq:Ee}
\mathbf{E^e}=\frac{1}{2}(\mathbf{C^e}-I), \qquad \mathbf{C^e}=\mathbf{F^{eT}}\mathbf{F^{e}}
\end{equation}
where $\mathbf{C}^e$ is the elastic right Cauchy-Green tensor defined in the intermediate configuration. 

\subsubsection{Thermoelasticity}\label{sec:cp:elasticity}
Under shock loading the stress response to the deviatoric and volumetric strain components is computed separately.
The deviatoric response is carried by the anisotropic crystal elasticity and the hydrostatic response is supplied by the equation of state. 
The second Piola-Kirchhoff stress in the intermediate configuration is
\begin{equation}\label{eq:Te}
    \mathbf{T^e} = \mathbb{C}:\mathbf{E^e}
\end{equation}
where $\mathbb{C}$ is the fourth-order elasticity tensor. 
For cubic symmetry $\mathbb{C}$ is parameterized by $C_{11}$, $C_{12}$ and $C_{44}$ and $\mathbf{R}\in SO(3)$, determined by the local lattice orientation.
Moreover, the elastic constants depend linearly on temperature and pressure \cite{lee2026bayesian}.
\begin{equation}\label{eq:CijTp}
    {C_{ij}(\theta, p) = C_{ij} + C_{ij,\theta}\left(\theta - \theta_\mathrm{C}\right) + C_{ij,p}\,p,}
\end{equation}
where $C_{ij}$ are the constants at reference temperature $\theta_c$ and reference pressure.
$ C_{ij,p}$ and $C_{ij,\theta}$ are the pressure and temperature derivatives of the elastic constants and $p$ is the pressure of the equation of state.
The corresponding Cauchy stress is
\begin{equation}\label{eq:Tcpl}
    \mathbf{T_\mathrm{cp}} = J^{-1}\mathbf{F^e}\mathbf{T^e}\mathbf{F^{eT}}, \qquad J=\det \mathbf{F}
\end{equation}
The resolved shear stress that drives slip on the slip system $\alpha$ is
\begin{equation}\label{eq:rss}
    \tau^{\alpha} = (\mathbf{C^e}\mathbf{T^e}):\mathbf{S_0^\alpha}
\end{equation}

\subsubsection{Dislocation interactions and slip resistances}\label{sec:cp:interactions}
The dislocation density $\rho^\alpha$ is the line length of dislocations of slip system $\alpha$ per unit volume in the intermediate configuration.
Slip system interactions are described through geometric projections of the total dislocation density onto each system \cite{ma2004, luscher2017, nguyen2021dynamic}. 
A dislocation gliding on a system $\alpha$ meets two kinds of neighboring dislocations.
Forest dislocations thread its glide plane and are long-range athermal obstacles.
Coplanar dislocations lie on parallel planes and act as short-range barriers that are overcome by thermal activation \cite{lloyd2014shock, nguyen2021dynamic}.
The two densities are obtained by projecting the total dislocation density of every slip system onto the plane normal $\mathbf{m}_0^\alpha$  and into the plane,
\begin{align}
  \varrho_\mathrm{for}^\alpha &= \sum_{\beta=1}^\mathrm{{n_{slip}}}\mathbf{A}_\mathrm{for}^{\alpha \beta} \varrho ^\beta,
  & \mathbf{A}_\mathrm{for}^{\alpha \beta} &= \frac{1}{2}\left | \mathbf{m}_0^{\alpha} \cdot \mathbf{s}_0^{\beta}   \right | + \frac{1}{2} \left |  \mathbf{m}_0^{\alpha} \cdot ( \mathbf{m}_0^{\beta}\times \mathbf{s}_0^{\beta} )  \right |   ,
    \label{eq:rhofor}\\
  \varrho_\mathrm{cop}^\alpha &= \sum_{\beta=1}^\mathrm{{n_{slip}}}\mathbf{A}_\mathrm{cop}^{\alpha \beta} \varrho ^\beta,
  & \mathbf{A}_\mathrm{cop}^{\alpha \beta} &= \frac{1}{2}\left\| \mathbf{m}_0^{\alpha} \times \mathbf{s}_0^{\beta}   \right\| + \frac{1}{2} \left\|  \mathbf{m}_0^{\alpha} \times ( \mathbf{m}_0^{\beta}\times \mathbf{s}_0^{\beta} )  \right\|  .
    \label{eq:rhocop}
\end{align}
Here, $A_\mathrm{for}^{\alpha\beta}$ and $A_\mathrm{cop}^{\alpha\beta}$ are the projection coefficients of the density of slip system $\beta$ onto the forest and coplanar populations of slip system $\alpha$.
Each coefficient is the sum of an edge and screw contribution with an equal mixture of the two assumed on every system \cite{luscher2017, nguyen2021dynamic}.
The athermal resistance $\tau_a^\alpha$ and the thermal barrier strength $\tau_r^\alpha$ are given by,
\begin{align}
  \tau_a^\alpha &= \tau_{a,0} + C_\mathrm{for}b\mu \sqrt{\varrho_\mathrm{for}^\alpha} ,
  &
    \tau_r^\alpha &= \tau_{r,0} + C_\mathrm{cop}b\mu \sqrt{\varrho_\mathrm{cop}^\alpha} ,
  \label{eq:tau}
\end{align}
where $\tau_{a,0}$ is the reference athermal resistance, $\tau_{r,0}^\alpha$ is the lattice (Peierls) resistance of the family of systems $\alpha$, $b$ is the magnitude of the Burgers vector, $\mu$
is the isotropic shear modulus, and $C_\mathrm{for}$ and $C_\mathrm{cop}$ are hardening coefficients.
The Peierls resistance takes one value on the $\{110\}$ and two values on the $\{112\}$ systems for the twinning and anti twinning sense of $\tau^\alpha$ \cite{nguyen2021dynamic}.

\subsubsection{Flow rule}\label{sec:cp:flowrule}
Total dislocation on each system is split into mobile and immobile fractions,
\begin{equation}\label{eq:rhosplit}
    \rho^{\alpha} = \rho_M^{\alpha} + \rho_I^{\alpha}
\end{equation}
and only the mobile fraction produces plastic slip.
The slip rate follows from the Orowan relation \cite{orowan1940,nguyen2021dynamic},
\begin{equation}\label{eq:orowan}
    \dot{\gamma}_p^\alpha = b \rho_M^\alpha v^\alpha
\end{equation}
where $v^\alpha$ is the signed mean glide velocity.
The mean glide velocity is obtained by treating glide as a sequence of waiting and running events between obstacles \cite{kocks1975, austin2011, nguyen2021dynamic}.
It is defined as,
\begin{equation}\label{eq:vel}
    v^\alpha = \frac{\bar L ^\alpha}{t_w^\alpha + t_r^\alpha} \operatorname{sign} (\tau ^\alpha), \qquad \bar{L^\alpha} = \frac{1}{\sqrt{\rho^\alpha_{cop}}}
\end{equation}
where $\bar{L}^\alpha$ is the mean spacing of coplanar obstacles, $t_w^\alpha$ is the time spent waiting at an obstacle, and $t_r^\alpha$ is the time spent running between obstacles against phonon drag. 
The waiting time follows an Arrhenius form with a Kocks-Argon-Ashby activation enthalpy \cite{kocks1975},
\begin{equation}\label{eq:tw}
    t_w^\alpha = \frac{1}{f_D} \left( \exp\! \left[  \frac{\Delta G (\tau^{\star \alpha })}{k_B \theta}  \right]  -1 \right), \qquad \Delta G (\tau^{\star \alpha}) = \Delta G_0 \left \langle 1 - \left( \frac{\tau^{\star \alpha}}{\tau_r ^\alpha} \right)^p   \right \rangle ^q
\end{equation}
where $f_D$ is the attempt frequency, $\Delta G_0$ is the zero-stress kink-pair formation energy, $k_B$ is the Boltzmann constant, and $p\in(0,1]$ and $q\in[1,2]$ control the shape of the obstacle profile. 
The effective stress available for barrier crossing is the overstress above the athermal resistance,
\begin{equation}\label{eq:taustar}
    \tau ^{\star \alpha} = \left \langle \left | \tau ^{\alpha}  \right | - \tau_a ^\alpha  \right \rangle
\end{equation}
where $\langle \cdot \rangle = \tfrac{1}{2}\left(|\cdot| + (\cdot)\right)$ denotes the Macaulay bracket. 
The running time is set by the drag-limited velocity between obstacles \cite{austin2011},
\begin{equation}\label{eq:vr}
    v_r^\alpha = \frac{\bar L^\alpha}{t_r^\alpha} =\frac{b \tau ^{\star \alpha}}{B^\alpha_{\star}}, \qquad B^\alpha_{\star}=\frac{B_0}{1-\left (\frac{v^\alpha_r}{c_s}\right)^2}
\end{equation}
where $B_0$ is the drag coefficient in the low-velocity limit, $c_s$ is the shear wave speed and $B_\star^\alpha$ is the velocity dependent drag coefficient.
\Cref{eq:vr} is implicit in $v_r^\alpha$ but admits the closed form
\begin{equation}\label{eq:vrclosed}
    v_r^\alpha = \frac{c_s^2}{2w^\alpha}\left (  \sqrt{1+\frac{4(w^\alpha)^2 }{c_s^2}} -1   \right), \qquad w^\alpha \stackrel{\textbf{def}}=\frac{b\tau^{\star \alpha}}{B_0}
\end{equation}
obtained as the positive root of $(w^\alpha/c_s^2)(v_r^\alpha)^2 + v_r^\alpha - w^\alpha = 0$. 
This recovers the Newtonian limit $v_r^\alpha \to w^\alpha$ for $w^\alpha \ll c_s$ and saturates at $v_r^\alpha \to c_s$ for $w^\alpha \gg c_s$.
\subsubsection{Dislocation density evolution}\label{sec:cp:evolution}
The mobile density evolves through multiplication, mutual annihilation, and trapping at obstacles. 
Trapped segments evolve the immobile density, which is depleted by dynamic recovery,
\begin{align}
\dot{\rho^{\alpha}_M}=\dot{\rho^{\alpha}_{mult}} -\dot{\rho^{\alpha}_{ann}} - \dot{\rho^{\alpha}_{trap}} ,
    \label{eq:rhoMdot}\\
\dot{\rho^{\alpha}_I}=\dot{\rho^{\alpha}_{trap}} -\dot{\rho^{\alpha}_{rec}} \approx f_{rec}\dot{\rho^{\alpha}_{trap}} ,
    \label{eq:rhoIdot}
\end{align}
where $f_{\text{rec}}^\alpha$ is the retained fraction of the trapped density.
Each contribution is proportional to the swept area rate $|v^\alpha|$ per unit mobile density \cite{austin2011, luscher2017, nguyen2021dynamic},
\begin{align}
  \dot{\rho^{\alpha}_{mult}}&=C_M \sqrt{\varrho_{for}^{\alpha}}\rho_M^\alpha \left |  v_{\alpha} \right |,
  &
    \dot{\rho^{\alpha}_{ann}}&=\frac{1}{4}C_A d_A^{\alpha} (\rho_M^{\alpha})^2 \left | v^{\alpha}\right | ,
  &
    \dot{\rho^{\alpha}_{trap}}&=C_T \sqrt{\varrho_{for}^{\alpha}}\rho_M^\alpha \left |  v_{\alpha} \right |,
    \label{eq:mult_ann_trap}
\end{align}
with $C_M$, $C_A$ and $C_T$ the multiplication, annihilation and trapping coefficients and $d_A^\alpha$ the annihilation capture radius. 
Rather than treating $C_A$ and $d_A^\alpha$ as independent, they are constrained by requiring that \cref{eq:rhoMdot}
admit a stationary point at a saturation density
$\rho_{M,\text{sat}}^\alpha$, which gives
\begin{equation}\label{eq:CAdA}
    C_A = 4 (C_M - C_T), \qquad d_A^{\alpha} =\frac{\sqrt{\varrho_{for}^\alpha}}{\rho_{sat,M}^{\alpha}}, \qquad \rho_{M,sat}^{\alpha}=\varsigma\rho_{I,sat}^{\alpha}
\end{equation}
where $\varsigma$ is the ratio of the mobile to the immobile saturation density.
The retained fraction is prescribed so that the net storage of immobile density vanishes at saturation and reverses above it \cite{nguyen2021dynamic},
\begin{equation}\label{eq:frec}
    f_{rec}^{\alpha} = 1 - \left ( \frac{\rho_I^{\alpha}}{\rho_{I,sat}^{\alpha}}  \right )^{n_\text{rec}}
\end{equation}
with $n_{\text{rec}}$ the inverse saturation exponent.
The saturation density is rate and temperature dependent, following the Kocks-Mecking type scaling
\begin{equation}\label{eq:rhoIsat}
 \rho_{I,sat}^{\alpha} = \max \left[\rho_{0,sat}\left ( \frac{|\dot{\gamma}_\alpha|}{\dot{\gamma}_0} \right )^{k_B\theta/A}, \rho_{I,0}^\alpha \right]
\end{equation}
where $\rho_{0,\text{sat}}$ is a reference saturation density, $\dot{\gamma}_0$ is a reference slip rate, and $A$ is a characteristic activation energy for recovery. 
The lower bound $\rho_{I,0}^\alpha$ is the initial immobile density which prevents depletion of the stored density below its annealed value.

\begin{table}[pos=!htbp]
\centering
\caption{Summary of model parameters. Type: L, literature or experiment. F, fitted in this work to the $\SI{196}{\meter\per\second}$ FSV of \cite{jones2018effect}. N, numerical or modeling choice. D, derived from other parameters.}
\label{tab:params}
\scriptsize
\renewcommand{\arraystretch}{0.98}
\begin{tabularx}{\linewidth}{>{$}l<{$} X r l c l}
\toprule
\multicolumn{1}{l}{Symbol} & Description & Value & Unit & Type & Ref\\
\midrule 
\multicolumn{6}{l}{\textit{\nameref{sec:equation_of_state} (\cref{sec:equation_of_state})}} \\
\rho_0 & Reference density & 16580 & kg\,m$^{-3}$ & L & \cite{jones2018effect} \\
C_0 & Bulk sound speed & 3430 & m\,s$^{-1}$ & L & \cite{jones2018effect} \\
s & Hugoniot slope & 1.19 & - & L & \cite{jones2018effect} \\
\Gamma_0 & Gr\"uneisen parameter & 1.67 & - & L & \cite{jones2018effect} \\
\midrule
\multicolumn{6}{l}{\textit{\nameref{sec:shock_av} (\cref{sec:shock_av})}} \\
C_Q & Quadratic viscosity coefficient &2.0 & - & N & \cite{wilkins1980} \\
C_L & Linear viscosity coefficient & 0.15 & - & N & \cite{wilkins1980} \\
\midrule
\multicolumn{6}{l}{\textit{\nameref{sec:cp:elasticity} (\cref{sec:cp:elasticity})}} \\
C_{11} & Cubic elastic constant at $\theta_\mathrm{C}$ & $261.0$ & GPa & L & \cite{nguyen2021dynamic}\\
C_{12} & Cubic elastic constant at $\theta_\mathrm{C}$ & $156.3$ & GPa & L & \cite{nguyen2021dynamic}\\
C_{44} & Cubic elastic constant at $\theta_\mathrm{C}$ & $82.60$ & GPa & L & \cite{nguyen2021dynamic}\\
C_{11,\theta} & Temperature derivative of $C_{11}$ & $-0.0245$ & GPa\,K$^{-1}$ & L & \cite{nguyen2021dynamic}\\
C_{12,\theta} & Temperature derivative of $C_{12}$ & $-0.0118$ & GPa\,K$^{-1}$ & L & \cite{nguyen2021dynamic}\\
C_{44,\theta} & Temperature derivative of $C_{44}$ & $-0.0149$ & GPa\,K$^{-1}$ & L & \cite{nguyen2021dynamic}\\
C_{11,p} & Pressure derivative of $C_{11}$ & $5.10$ & - & L & \cite{nguyen2021dynamic}\\
C_{12,p} & Pressure derivative of $C_{12}$ & $3.14$ & - & L & \cite{nguyen2021dynamic}\\
C_{44,p} & Pressure derivative of $C_{44}$ & $0.995$ & - & L & \cite{nguyen2021dynamic}\\
\theta_\mathrm{C} & Reference temperature of the elastic constants & $300$ & K & L & \cite{nguyen2021dynamic}\\
\mu & Isotropic shear modulus & $62.43$ & GPa & L & \cite{nguyen2021dynamic}\\
\midrule
\multicolumn{6}{l}{\textit{Slip resistance and glide kinetics (\cref{sec:cp:interactions,sec:cp:flowrule})}} \\
b & Burgers vector magnitude & $2.863\times10^{-10}$ & m & L & \cite{nguyen2021dynamic}\\
\tau_{a,0} & Reference athermal resistance & $27$ & MPa & L & \cite{nguyen2021dynamic}\\
C_\mathrm{for} & Forest hardening coefficient & $0$ & - & L & \cite{nguyen2021dynamic}\\
C_\mathrm{cop} & Coplanar hardening coefficient & $0.5$ & - & L & \cite{nguyen2021dynamic}\\
\tau_{r,0}^{110} & Peierls resistance, $\{110\}$ & $655$ & MPa & L & \cite{nguyen2021dynamic}\\
\tau_{r,0}^{112,\mathrm{T}} & Peierls resistance, $\{112\}$, twinning sense & $575$ & MPa & L & \cite{nguyen2021dynamic}\\
\tau_{r,0}^{112,\mathrm{AT}} & Peierls resistance, $\{112\}$, antitwinning sense & $1075$ & MPa & L & \cite{nguyen2021dynamic}\\
g_0 & Activation energy coefficient & $0.2$ & - & L & \cite{nguyen2021dynamic}\\
\Delta G_0 & Activation energy at zero stress, $g_0\,\mu\, b^3$ & $2.93\times10^{-19}$ & J & D & \cite{nguyen2021dynamic}\\
p & Obstacle profile exponent & $0.3387$ & - & L & \cite{nguyen2021dynamic}\\
q & Obstacle profile exponent & $1.9519$ & - & L & \cite{nguyen2021dynamic}\\
f_D & Attempt frequency & $1.0\times10^{11}$ & s$^{-1}$ & L & \cite{nguyen2021dynamic}\\
B_0 & Drag coefficient & $2.9574\times10^{-5}$ & Pa\,s & L & \cite{nguyen2021dynamic}\\
c_s & Shear wave speed & $1937$ & m\,s$^{-1}$ & L & \cite{nguyen2021dynamic}\\
\midrule
\multicolumn{6}{l}{\textit{Dislocation density evolution (\cref{sec:cp:evolution})}} \\
\textstyle\sum_\alpha \rho^\alpha_0 & Initial total dislocation density & $6.0\times10^{12}$ & m$^{-2}$ & N & this work\\
\rho_{M,0}^\alpha/\rho_0^\alpha & Initial mobile fraction & 0.0025 & - & N & \cite{nguyen2021dynamic}\\
C_M & Multiplication coefficient & $0.3502$ & - & L & \cite{nguyen2021dynamic}\\
C_T/C_M & Trapping-to-multiplication ratio & $0.8836$ & - & L & \cite{nguyen2021dynamic}\\
C_A & Annihilation coefficient, $4(C_M-C_T)$ & $0.1631$ & - & D & \cite{nguyen2021dynamic}\\
\varsigma & Mobile to immobile saturation ratio & $0.1$ & - & L & \cite{nguyen2021dynamic}\\
\rho_{0,\mathrm{sat}} & Reference saturation density & $2.6596\times10^{15}$ & m$^{-2}$ & L & \cite{nguyen2021dynamic}\\
\dot{\gamma}_0 & Reference slip rate & $2.5143\times10^{9}$ & s$^{-1}$ & L & \cite{nguyen2021dynamic}\\
A & Activation energy for recovery & $1.5762\times10^{-19}$ & J & L & \cite{nguyen2021dynamic}\\
n_\mathrm{rec} & Inverse saturation exponent & $4$ & - & L & \cite{nguyen2021dynamic}\\
\midrule
\multicolumn{6}{l}{\textit{Thermal coupling (\cref{sec:coupling})}} \\
c_v & Specific heat at constant volume & $140$ & J\,kg$^{-1}$\,K$^{-1}$ & L & \cite{nguyen2021dynamic}\\
\theta_0 & Initial temperature & $294$ & K & L & \cite{jones2018effect}\\
 \midrule
\multicolumn{6}{l}{\textit{Continuum damage model (\cref{sec:damage})}} \\
w_c & Critical energy density & $2.0\times10^{8}$ & J\,m$^{-3}$ & N & \cite{pham2011}\\
G_c & Critical energy release rate, $c_w w_c \ell$ & $3.0\times10^{3}$ & J\,m$^{-2}$ & D & \cite{pham2011}\\
Y_f & Flow stress of the failure criterion & 1.0 & GPa & N & this work\\
\varepsilon_f & Failure strain & 0.6 & - & F & this work\\
W_c & Critical plastic work, $Y_f\varepsilon_f$ & $6.0\times10^{8}$ & J\,m$^{-3}$ & D & \cite{borden2016,miehe2016}\\
\zeta & Slope of the driving state function & 10 & - & N & \cite{miehe2015}\\
\ell & Regularization length & 5.7 & \textmu m & N & \cite{pham2011}\\
M & Mobility & 1.7 & m$^{3}$\,J$^{-1}$\,s$^{-1}$ & D & \cite{miehe2010}\\
\midrule
\multicolumn{6}{l}{\textit{Void nucleation criterion (\cref{sec:nucleation_model})}} \\
g_1 & Microstructural barrier coefficient & 3.0 & - & N & this work\\
h_\mathrm{ref} & Reference susceptibility & 0.39 & - & N & this work\\
h_\mathrm{min} & Susceptibility threshold & 0.03 & - & N & this work\\
m & Stress exponent & 2 & - & L & \cite{fisher1948fracture}\\
\sigma_c & Onset stress & 5.80 & GPa & F & this work\\
\mathbf{B} & Mechanical barrier at $\sigma_c$, \cref{eq:onset} & 10 & - & F & this work\\
\sigma_s & Stress scale, $\sigma_c\mathbf{B}^{1/m}$, \cref{eq:onset} & 18.3 & GPa & D & this work\\
t_s & Start time of the nucleation law & 1.645 & \textmu s & F & this work\\
\theta_\mathrm{ref} & Reference temperature & 294 & K & N & this work\\
\bottomrule
\end{tabularx}
\end{table}

\subsubsection{Hydrodynamic coupling}\label{sec:coupling}

This section presents the coupling between the shock physics and crystal plasticity.
Given an EOS pressure from the density and the internal energy, the crystal plasticity model returns the full stress tensor from the deformation gradient. 
Only one stress can enter the momentum balance of \cref{eq:conservation_eqn} and only one temperature can enter the flow rule of \cref{eq:tw}.
The coupling therefore has two parts.
The decomposition of the stress between the shock physics and the crystal plasticity model is given first and the exchange of plastic work and temperature between them is given second.

The stress response to the deviatoric and volumetric parts of the deformation is computed separately, following \cite{lee2026bayesian}.
The deviatoric part of the Cauchy stress comes from the crystal plasticity model \cref{eq:Te,eq:Tcpl}, and the hydrostatic part comes from the Mie-Gr\"uneisen equation of state \cref{eq:mg}. 
Thus, the coupled Cauchy stress is,
\begin{equation}\label{eq:Tcorrect}
    \mathbf{T} = \mathbf{T_\mathrm{cp}}-(p_\mathrm{eos} -p_\mathrm{cp} )\mathbf{I} - \mathbf{q}, \qquad p_\mathrm{cpl} = -\frac{1}{3} \operatorname{tr}\mathbf{T}_\mathrm{cp}
\end{equation}
where $\mathbf{T}_\mathrm{cp}$ is the Cauchy stress of the crystal plasticity model, $p_\mathrm{cp}$ is the mean pressure calculated from it, $p_\mathrm{eos}$ is the pressure of the equation of state and $\mathbf{q}$ is the artificial viscous stress from \cref{eq:av2d}.
The first Piola-Kirchhoff stress that enters \cref{eq:conservation_eqn} is  $\mathbf{P} = J\mathbf{T}\mathbf{F}^{-T}$.
While the stress of the intact material is described by \cref{eq:Tcorrect}, its degradation in the presence of damage is explained in \cref{sec:damage}.
Since the Schmid tensors in the intermediate and the deformed configurations are traceless,
\begin{equation}\label{eq:traceless}
    \mathbf{I}:\mathbf{S}^{\alpha} = \operatorname{tr}\left (  \mathbf{F}^e \mathbf{S}_0^{\alpha} \mathbf{F}^{e-1} \right ) = \operatorname{tr} \mathbf{S_0^\alpha} = \mathbf{s_0^\alpha}\cdot \mathbf{m_0^\alpha} = 0
\end{equation}
the pressure correction does not influence the slip rate.
The pressure of the equation of state enters the crystal plasticity model only through the pressure dependence of the elastic constants in \cref{eq:CijTp}, and for this purpose the pressure of the previous stage of the time step is used.

The internal energy $e$ receives all the work done on the material \cite{wilkins1964}.
In a shock physics formulation, the thermal energy is the excess of $e$ over the reference energy $e_\mathrm{ref}$ of \cref{eq:pref} \cite{meyers1994dynamic, clayton2019}.
With the crystal plasticity model coupled with the shock physics, $e$ also holds the deviatoric elastic energy $\psi_\mathrm{dev}$ and the energy stored in dislocations $\psi_s$.
So the thermal energy per unit reference volume and pressure is
\begin{equation}\label{eq:mgtension}
  \varepsilon_\mathrm{th} = \rho_0\left(e - e_\mathrm{ref}\right) - \psi_\mathrm{dev} - \psi_s,
  \qquad
  p = P_\mathrm{ref}(\eta) + \Gamma_0\,\varepsilon_\mathrm{th}
\end{equation}
where $\psi_\mathrm{dev}$ is the elastic energy of the crystal, and $\psi_s$ is the energy of the dislocations,
\begin{equation}\label{eq:psis}
  \psi_{s} = \tfrac{1}{2}\,\mu b^{2}\sum_{\alpha}\left(\rho^{\alpha}_{M} + \rho^{\alpha}_{I}\right) - \psi_{s,0}
\end{equation}
where $\mu$ is the shear modulus, $b$ is the magnitude of the Burgers vector, $\rho^\alpha_M$ and $\rho^\alpha_I$ are the mobile and immobile dislocation densities of slip system $\alpha$, and $\psi_{s,0}$ is the value at the initial densities \cite{luscher2017, nguyen2021dynamic}.
The heat of plastic work is $\dot q_p = \dot w_p - \dot\psi_s$, where $\dot w_p = \sum_\alpha \tau^\alpha \dot\gamma^\alpha$ is the plastic work rate of the crystal plasticity model.
The temperature evolves as
\begin{equation}\label{eq:theta}
  \rho_0 c_v\,\dot\theta = -\rho_0 c_v\,\Gamma_0\,\theta\,\dot J + \dot q_p + \dot q_\mathrm{av} + \dot q_d
\end{equation}
where $c_v$ is the specific heat and $\dot q_\mathrm{av}$ is the heat of the artificial viscosity \cite{meyers1994dynamic, nguyen2021dynamic}.
The heat released during void formation, $\dot q_d$ is the heat released from damage growth and is described in \cref{sec:damage}.
The evolved temperature enters the waiting time of \cref{eq:tw}, the elastic constants of \cref{eq:CijTp}, and the saturation density of \cref{eq:rhoIsat}.

\subsection{Continuum damage model}\label{sec:damage} 
The crystal plasticity model of \cref{sec:crystalplasticity} resolves the grain scale fields that drive void formation and growth but does not itself evolve damage.
This section describes the continuum damage model that uses those fields and returns a spatially resolved damage variable that represents the local volume fraction of the void phase. 
The formulation is the AT1 regularization of the variational approach to fracture \cite{bourdin2000,pham2011}.
AT1 is adopted in preference to AT2 because it possesses a genuine elastic threshold \cite{pham2011}.
Below a critical stored energy the damage field remains zero, so the material is not progressively softened during compression and the onset of damage is set by a material parameter rather than by the regularization.

Damage is represented by a continuous order parameter $d(\mathbf{x},t)\in[0,1]$, with $d=0$ for intact material and $d=1$ for a fully formed void. 
The total free energy per unit reference volume is
\begin{equation}\label{eq:dmenergy}
      \Psi = g(d)\left[\psi^{+}_{\mathrm{vol}} + \psi_{\mathrm{dev}} + \psi_{s}\right]       + \psi^{-}_{\mathrm{vol}}  +w_c[w(d) + \ell^2|\nabla d|^2],
\end{equation}
where $\psi^{\pm}_{\mathrm{vol}}$ are the volumetric elastic energies in tension and compression, $\psi_{\mathrm{dev}}$ is the deviatoric elastic energy of the crystal, $\psi_s$ is the energy stored in dislocations, $w_c$ is the critical energy density, and $\ell$ is the regularization length.
The critical energy release rate is $G_c=c_w w_c \ell$ \cite{pham2011}.
The first two terms are the degraded and undegraded parts of the elastic energy in the volumetric split of the variational approach to fracture \cite{amor2009}.
The energy stored in dislocations is taken analogous to the energy of cold work and is placed inside the degraded bracket.
Energy from cold work is degraded in the phase field models of ductile fracture at finite strains \cite{miehe2016}.
The last term is the regularized fracture energy \cite{bourdin2000, pham2011}.
The degradation function and the AT1 dissipation potential are
\begin{equation}\label{eq:gW}
  g(d) = (1-k_r)(1-d)^{2} + k_r,
  \qquad
  w(d) = d,
  \qquad
  c_w = 4\!\int_{0}^{1}\!\sqrt{w(s)}\,\mathrm{d}s = \tfrac{8}{3}
\end{equation}
where $k_r$ is a residual that keeps the degraded response non-singular \cite{bourdin2000}.
The quadratic degradation function is the standard choice of the variational approach \cite{bourdin2000, miehe2010}.
The quantities $w_c$ and $\ell$ set the strength of the intact material,
$\sqrt{\rho_0 C_0^2 w_c}$, in uniaxial tension and the width of the damage band, $2 \ell$ \cite{pham2011}. 

In \cref{eq:dmenergy}, the degradation function $g(d)$ degrades the elastic energy in tension and the energy stored in dislocations.
These two energies drop as damage grows.
The elastic energy in compression is not degraded, so damaged material still carries full compression \cite{amor2009, miehe2016}.
Since slip does not change volume, the volume ratio of the total deformation is the same as that of the elastic deformation, $J=\det\mathbf{F}=\det\mathbf{F}^{e}$ \cite{nguyen2021dynamic}.
The volumetric energy is split into a tensile part for $J>1$ and a compressive part for $J\le1$ \cite{amor2009, borden2016}.
The tensile part is taken from the equation of state and not from the crystal plasticity, since the equation of state sets the mean stress in \cref{eq:Tcorrect}.
The two parts are
\begin{equation}\label{eq:psiplus}
  \psi^{+}_{\mathrm{vol}} = \begin{cases} \tfrac{1}{2}\,\rho_0 C_0^{2}\,(J-1)^{2} & J > 1, \\ 0 & J \le 1, \end{cases}
  \qquad
  \psi^{-}_{\mathrm{vol}} = \begin{cases} 0 & J > 1, \\ \rho_0\,e_{\mathrm{ref}}(\eta) & J \le 1, \end{cases}
\end{equation}
where $\rho_0 C_0^2$ is the bulk modulus of the tensile branch of \cref{eq:pref}, $e_{\mathrm{ref}}(\eta)$ is the reference energy of \cref{eq:pref}, and $\eta = 1 - \rho_0/\rho = 1 - 1/J$ is the volumetric strain.
In tension the volumetric energy is that of the linear elastic branch and it is degraded.
In compression, it is the energy of the Hugoniot reference curve and it is not degraded.
This is the condition of the volumetric split, in which a closed crack still transmits compression \cite{amor2009}.

The deviatoric energy $\psi_{\mathrm{dev}}$ is the elastic energy of the crystal of \cref{eq:Te}.
It is computed on the isochoric part of $\mathbf{F}^{e}$ with the undamaged constants \cite{borden2016}.
The stored energy $\psi_s$ is the line energy of the dislocations \cref{eq:psis} \cite{luscher2017, nguyen2021dynamic}.
These are the two energies that \cref{sec:coupling} subtracts from the internal energy before the equation of state is evaluated.

The variational derivative of \cref{eq:dmenergy} with respect to $d$ is
\begin{equation}\label{eq:Ypsi}
  Y_\Psi = 2(1-k_r)(1-d)\left[\psi^{+}_{\mathrm{vol}}+\psi_{\mathrm{dev}}+\psi_{s}\right]
          - w_c + 2\,w_c\,\ell^{2}\,\nabla^{2}d
\end{equation}
In the phase field model of brittle fracture this force drives the damage variable, and damage starts where the stored elastic energy exceeds $w_c/2$ \cite{bourdin2000, pham2011, miehe2010}.
Ductile fracture is driven by plastic flow and not by stored elastic energy.
The phase field models of ductile fracture therefore extend \cref{eq:Ypsi} in one of two ways. 

In the first approach, a plastic energy is included in the free energy and degraded by $g(d)$ and the damage force is still the derivative of the free energy \cite{alessi2015, miehe2016, borden2016}.
The stored energy in cold work is the part of the plastic energy that can be degraded without violating the dissipation inequality \cite{miehe2016}.
In the second approach, the energies in \cref{eq:Ypsi} are replaced by a driving state function.
This function is not derived from the free energy but from a failure criterion of the matrix \cite{miehe2015a,miehe2015,ambati2015}.
In the present approach, the second way is adopted with the plastic work as the failure criterion.
The free energy retains its two other roles.
It gives the stress of the damaged material that replaces \cref{eq:Tcorrect} in the momentum balance.
\begin{equation}\label{eq:Tdamaged}
  \mathbf{T} = g(d)\,\operatorname{dev}\mathbf{T}_\mathrm{cp} - p\,\mathbf{I} - g_q\,\mathbf{q},
  \qquad
  p = \begin{cases} g(d)\bigl[\rho_0C_0^2(1-J) + \Gamma_0\varepsilon_\mathrm{th}\bigr] & J>1,\\[2pt]
                    P_H(\eta) + \Gamma_0\varepsilon_\mathrm{th} & J\le1, \end{cases}
  \qquad
  g_q = \begin{cases} g(d) & J>1,\\ 1 & J\le1. \end{cases}
\end{equation}
Under tension, the deviatoric part of the Cauchy stress, and the entire pressure and the viscous part of the stress are completely degraded by $g(d)$.
Under compression, only the deviatoric part of the Cauchy stress is degraded but the compressive stress and viscous stress are undegraded and transmitted by the damaged material \cite{amor2009}.
The crystal plasticity model itself is undegraded.
It computes the undegraded $\mathbf{T}_\mathrm{cp}$, slip rates and dislocation densities without any reference to $d$.
Therefore, the undegraded stress carried by the intact material yields and hardens the crystal lattice \cite{lemaitre1985}.
However, $g(d)$ reduces the stress supplied to the momentum balance and the plastic work supplied to the energy balance.
This choice keeps the plastic work driving \cref{eq:Hdrv} active during void growth which would otherwise be suppressed by degradation \cite{borden2016}.
Once the material is completely damaged (d=1), it carries no deviatoric stress and no pressure for $J\ge1$.
It carries the pressure $k_r\rho_0C_0^2(1/J-1)$ for $J<1$.
Its crystal state, thermal energy and plastic work are reset to their initial values.

Second, the free energy also gives the energy that is released when damage grows, which is $-(\partial\psi/\partial d)\,\Delta d$ for an increment $\Delta d$.

The failure criterion is the plastic work done in tension relative to a critical value,
\begin{equation}\label{eq:Dtilde}
  \tilde{D} = \frac{W_p^{+}}{W_c},
  \qquad
  W_c = Y_f\,\varepsilon_f,
  \qquad
  \dot{W}_p^{+} = \begin{cases} \dot{w}_p & J > 1, \\ 0 & J \le 1, \end{cases}
\end{equation}
where $W_p^{+}$ is the plastic work accumulated while the material is under tension, $W_c$ is the critical plastic work, $Y_f$ is the flow stress, and $\varepsilon_f$ is the failure strain.
A critical plastic work as the failure criterion is taken from the ductile phase field damage models \cite{borden2016, miehe2016}.
In those models the plastic work is accumulated over the entire loading history but in the present work only the plastic work done while the material is under tension is included.
The critical work $W_c = Y_f\varepsilon_f$ is the work to failure of a tensile test.
Both $Y_f$ and $\varepsilon_f$ are taken as inputs to the simulation.
The flow stress $Y_f$ is taken as the representative flow stress of tantalum under shock loading, and the failure strain $\varepsilon_f$ is a calibrated parameter of the damage model.

The driving state function is
\begin{equation}\label{eq:Hdrv}
  \mathcal{H} = w_c\,\zeta\,\chi\,\bigl\langle\tilde{D}-1\bigr\rangle_{+},
  \qquad
  \chi = \begin{cases} 1 & \text{within } 2\ell \text{ of a void and } J > 1, \\ 0 & \text{otherwise}, \end{cases}
\end{equation}
where $\zeta$ is a dimensionless slope and $\langle\cdot\rangle_+$ is the Macaulay bracket.
The form $\zeta\langle\tilde{D}-1\rangle_{+}$ is the driving state function of the ductile phase field models \cite{miehe2015}.
The force is zero below the failure criterion and grows with the slope $\zeta$ above it and this slope is an input to the simulation.
The present form of the driving state function differs from the original form in three ways.
First, the original form is the sum of an elastic and a plastic part but the present form uses only the plastic part to drive damage.
Second, the original form ensures irreversibility by keeping the largest value of the driving force over the loading history but here it is enforced through the evolution law of \cref{eq:at1evolution}.
Third, the factor $\chi$ confines damage to dilated material only within $2\ell$ of an existing void \cite{pham2011}.

The driving force of the damage evolution is \cref{eq:Ypsi} with the energies replaced by $\mathcal{H}$.
It is given by
\begin{equation}\label{eq:drivingforce}
  Y = \underbrace{2(1-k_r)(1-d)\,\mathcal{H}}_{\text{mechanical}}
      \;\underbrace{-\;w_c}_{\text{threshold}}
      \;+\;\underbrace{2\,w_c\,\ell^{2}\,\nabla^{2}d}_{\text{gradient}}
\end{equation}
The threshold and the gradient terms are those of the AT1 model \cite{pham2011}.
The damage variable follows a rate dependent gradient flow \cite{miehe2010}.

\begin{equation}\label{eq:at1evolution}
  \dot{d} = M\,\langle Y\rangle_{+}
\end{equation}
where $M$ is the mobility.
The bracket keeps $\dot d \ge 0$, so that the damage cannot heal.
This is the viscous regularization of the damage variable with irreversibility enforced on the rate \cite{miehe2010}.
The mobility, $M$ is set by a relaxation time, $M=1/w_c\tau_d$ where $\tau_d = \ell/c_s$ so that the damage front relaxes over the time that the shear wave takes to cross $\ell$.

The growth of damage repartitions the energy between the thermal, stored and fractured parts but does not change the internal energy $e$ of \cref{eq:conservation_eqn}.
In a small damage increment of $\Delta d$ at fixed temperature and deformation, the energy released is,
\begin{equation}
    \Delta \psi_{rel} = [g(d) -g(d+\Delta d)](\psi_{vol}^+  +\psi_{dev} +\psi_s)
\end{equation}
and creates the fracture energy
\begin{equation}
    \Delta \psi_{frac} = w_c[1 - 2\ell^2\nabla^2d]\Delta d
\end{equation}
which is the work done against the threshold and the gradient terms of \cref{eq:drivingforce}.
When $\psi_{rel} \ge \psi_{frac}$, the fracture is paid from the released energy and the rest is dissipated as heat.
But if  $\psi_{frac} > \psi_{rel}$, then the deficit is paid from the plastic heat $g \dot q_p \Delta t$ of the current time step and later from the accumulated plastic heat from earlier steps so that the material cools down \cite{borden2016,miehe2015}.
The plastic work is therefore partitioned between the fracture energy and heat.
Therefore, the dissipation inequality $\sum g \Delta w_p  \ge \sum (\psi_{frac} -\psi_{rel})$ holds at all material points \cite{miehe2015}.
A deficit that exceeds the available plastic heat is negligible in simulations but is recorded separately.

The material that becomes a void either by insertion of a void or by growth of the initially inserted void leaves the energy balance.
Its stored energy, plastic work and heat are removed from the system but are tracked so that the energy balance of the remaining material can be checked.
Also, when a void is inserted in the domain, its surface energy $2\gamma$ per unit length is charged to the fracture energy.
It is nucleated only when the stored energy that the material releases when it is replaced by the void covers the surface energy.
Heat released during void growth, $\dot q_d$, enters the temperature evolution of \cref{eq:theta}.

\subsection{Void nucleation criterion}\label{sec:nucleation_model}
Damage in spall initiates through the formation and growth of voids.
Because void initiation begins at the atomistic level, a probabilistic model is necessary to determine the criteria favorable for void growth, which can then be used to insert pre-formed voids into the simulation.
The probability of nucleation can be considered as an Arrhenius rate form that occurs with probability $p_{nuc}$, given by the Boltzmann factor
\begin{equation}\label{eq:boltzmann}
    p_{nuc} = \exp \left(-\frac{E}{k_B \theta}\right),
\end{equation}
where $E$ is an activation energy, $k_B$ is the Boltzmann constant and $\theta$ is the temperature.
In the simulations conducted in this work, $\theta$ is held at $\theta_{ref}$.
Microstructure and boundary awareness is conveyed through the spatial dependence of $E$.
The activation energy also depends on the local stress state.
The following form of the activation energy, accounting for both of these effects, is proposed:
\begin{equation}\label{eq:nucleation}
     E ={k_B \theta_{ref}} \left[ g_1(h_{ref}-h)+ \left (\frac{\sigma_s}{\sigma_I} \right)^m \right]
\end{equation}
where $h_{ref},g_1,\sigma_l,m$ are model parameters, $\theta_{ref}$ is the reference temperature and $\sigma_I$ is the largest non negative principal value of the Cauchy stress.
Here, $h \in [0,1]$ is the susceptibility field.
It is a dimensionless measure of the likelihood of a grain boundary to nucleate a void.
It carries the dependence of the barrier on the microstructure and grain boundary properties.

The locations where $h<h_{min}$ are not nucleation sites and therefore the threshold $h_{min}$ confines nucleation to grain boundaries.
The second term is the mechanical barrier with $m=2$.
The value of $m$ is taken from the classical cavitation model\cite{fisher1948fracture}.
While $m=2$ is generally used for spherical voids, it has been retained in this work even though the domain of interest is two dimensional. 
The stress $\sigma_s$ sets the scale of the mechanical barrier.
At $\sigma_I = \sigma_s$, the barrier equals $k_B \theta$ and the barrier falls as $\sigma_I^{-2}$ above it.
It is related to the onset stress by
\begin{equation}\label{eq:onset}
    \sigma_s = \sigma_c \mathbf{B}^{1/m}
\end{equation}
where $\mathbf{B} = (\sigma_s/\sigma_c)^m$ is the mechanical barrier at $\sigma_I = \sigma_c$ in the units of $k_B\theta_{ref}$.
At the onset stress, a cell with $h=h_{ref}$ nucleates at a rate of $e^{-\mathbf{B}}$ per reference area $v_0$ over a reference time $t_0$.
The parameter $\mathbf{B}$ controls how fast the rate falls below $\sigma_c$.
Therefore a small value of $\mathbf{B}$ allows nucleation below $\sigma_c$ and a large value of $\mathbf{B}$ restricts nucleation events to be strictly above $\sigma_c$.
None of the parameters used in the nucleation model are borrowed from experimental measurements.
However, four different parameters are calibrated to capture the FSV for the $\SI{196}{\meter\per\second}$ experimental measurements.
The onset stress $\sigma_c$ and start time $t_s$ are calibrated to capture the pullback minimum.
The barrier height $\mathbf{B}$ and the rate scale $v_0t_0$ are fitted to capture the first reload.
This fitting is strictly restricted to the $\SI{196}{\meter\per\second}$ case and is validated against the $\SI{214}{\meter\per\second}$ experimental measurements.

The probability in \cref{eq:boltzmann} is the probability of the occurrence of a nucleation event in a reference area $v_0$ over a reference time $t_0$ \cite{gokuli2021multiphase}.
Only the product $v_0 t_0$ enters the model.
The nucleation law is evaluated at the finest grid level at every time step.
The probability that a nucleation event occurs in a cell of area $A_{i}$ during timestep $\Delta t$ follows from \cref{eq:boltzmann} as
\begin{equation}
    P_i = 1 - (1 - p_{nuc})^{A_i \Delta t / t_0 v_0} = 1 - \exp(-\lambda_i \Delta t), \qquad\lambda_i = \frac{A_i}{t_0 v_0}\,\frac{1}{N^2}\sum_{s=1}^{N^2} \left[-\ln\left(1 - p_{nuc}(h_s, \sigma_I, \theta)\right)\right],
\end{equation}
where $\lambda_i$ is the nucleation rate of a cell and $h_s$ is the void nucleation susceptibility field at the $N \times N$ sample points of a cell.
A sample point in which $h_s < h_{min}$ is not a nucleation site and its contribution is zero to the sum.
The total nucleation rate of a region is independent of cell size as long as the susceptibility field is resolved by the cell.
The probability $p_{nuc}$ is bounded by $1-10^{-12}$ so that the rate remains finite and $P_i$ lies in [0,1) by construction.

Each nucleation event is determined by a random threshold algorithm.
Each cell accumulates the quantity $K_i = \int_{}^{}\lambda_i dt$, which is integrated over the timesteps of the finest level via the trapezoidal rule.
Hence the probability of an event in a cell by time $t$ becomes $1 - \exp(-K_i)$, independent of the timestep.
A void nucleates in a cell $i$ at the first step in which $K_i > \zeta_i$ where $\zeta_i = -\ln U_i$ and $U_i$ is a random number in (0,1) drawn once per cell using a fixed seed.
If a cell undergoes compression, the rate is zero and $K_i$ remains at its current value.
Therefore, $K_i$ accumulated during tension is retained and continues to grow when the cell experiences tension again.
A cell can nucleate a void only once and a spacing factor, $r_{space} \ge (r_n + 2 \ell)$ is incorporated to make sure two nucleating voids do not touch each other.
It is important that nucleation statistics is grid and timestep independent, which is important when solving on an adaptively refined grid to avoid grid dependency.
The nucleation rate, $\lambda_i$ is dependent on the cell area $A_i$ and therefore, if a cell is divided into smaller cells, the rates of the smaller cells add up to the rate of the original cell, provided the smaller cells have the same $h$ and $\sigma_I$.
This ensures constant nucleation rate of a region across all grid levels.

The criterion therefore requires the field $h$ throughout the microstructure. 
In this work $h$ is predicted from grain orientations and grain boundary energies by employing computer vision. 

\section{Computer vision model to predict nucleation with microstructure and grain boundary awareness}\label{sec:machine_learning}

\begin{figure}
  \centering
  \begin{minipage}[c]{0.12\linewidth}
    \raggedleft
    \textbf{Micrograph}\\[1.2cm]
    \textbf{Repaired \\microstructure}\\[1.2cm]
     \textbf{GBE}\\[1.2cm]
    \textbf{Ground truth}\\[1.2cm]
    \textbf{Prediction}
  \end{minipage}%
  \hfill
  \begin{minipage}[c]{0.87\linewidth}
    \includegraphics[width=\textwidth]
{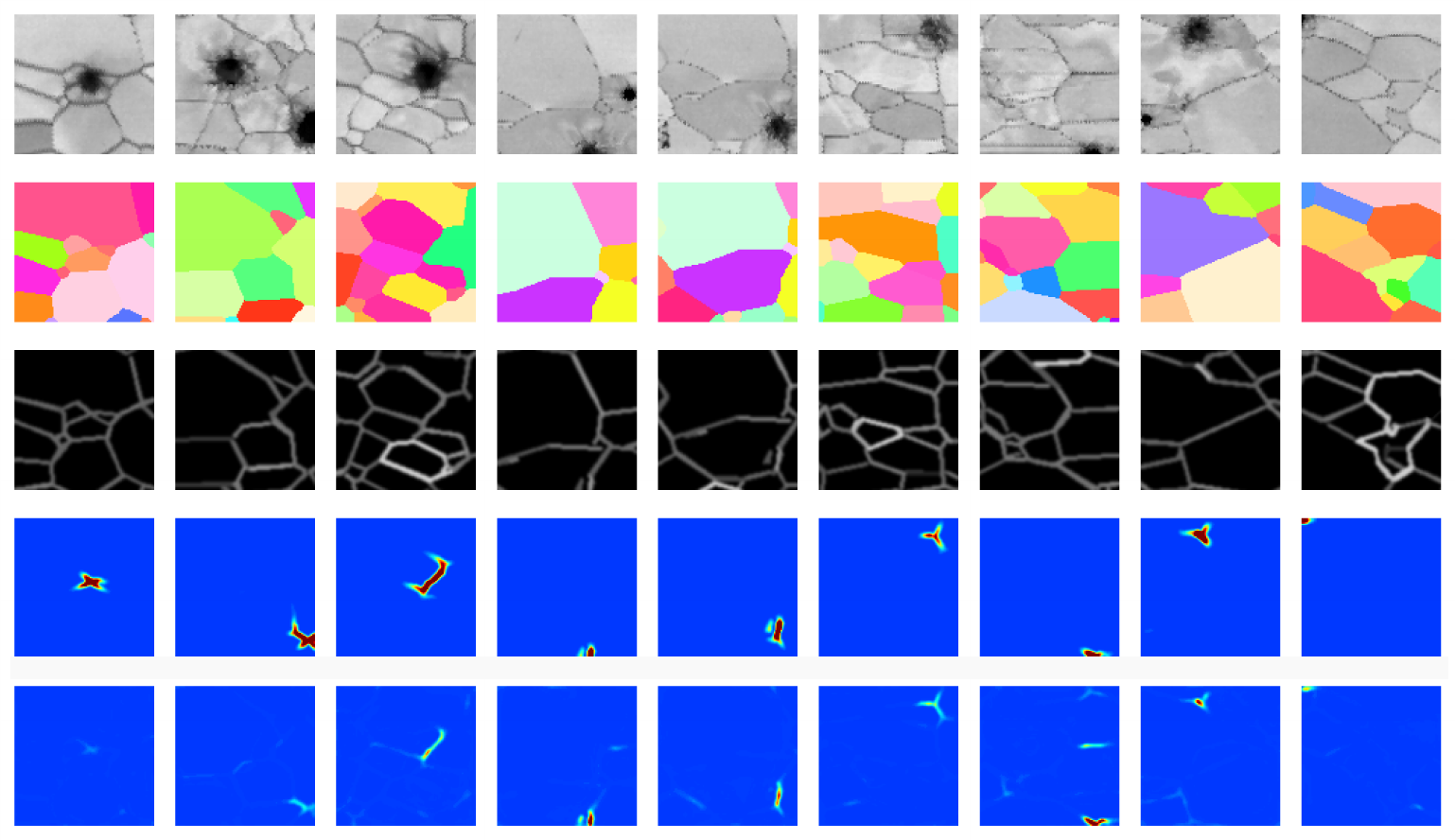}
  \end{minipage}
  \caption{
  A representative selection of the ML model predictions on the testing set with high PCC values (PCC $>$ 0.8) for nine mini-microstructures. 
  Each column consists of a post-mortem micrograph (row 1), reconstructed microstructure (row 2), grain boundary energy (row 3) the ground truth nucleation susceptibility field (row 4) and the actual prediction made by the machine learning model (row 5).
  The micrograph is only included for reference.
  The machine learning model receives the reconstructed microstructure which encodes grain orientation, and the GBE to predict void nucleation susceptibility. 
  The "ground truth" is generated by performing a pixelwise multiplication between the segmented void map generated from the micrograph and the GBE field and is visible to the model only via the loss function during training. 
  Colors correspond to no void nucleation likelihood (0, blue) and high void nucleation likelihood (1, red), with the intensity values clamped between 0.05 and 0.6 for better visualization. 
  }
  \label{fig:ml_test}
\end{figure}

It is well known that void nucleation depends on microstructural features and grain boundary properties. 
Experimental methods and molecular dynamics simulations have been developed to correlate microstructural features with void nucleation but they fail to build a correlation between a single microstructural feature and void nucleation \cite{fensin2014some,chen2020associating,nguyen2019role}.
But recently developed ML models are able to predict void nucleation from microstructural features and GBE.
The current work incorporates the stress from the crystal plasticity model, as well as the void nucleation predictions from the ML model to nucleate voids in simulations.
A brief description of the ML model has been presented here but the reader is referred to \cite{anantharanga2026predicting} for a complete description.

The training data for the ML model is drawn from post incipient spall micrographs obtained from symmetric plate impact experiments \cite{jones2018effect}  (\cref{fig:ml_test}, row 1).
While the voids in the micrographs are easily visible, the features leading to their nucleation exist in pristine microstructures, and therefore the ML model must be exclusively based on features in the microstructure prior to damage.
Since the images of pristine micrographs are unavailable, a microstructure reconstruction technique is incorporated to recover an estimate of the pristine state.
Regions in which extensive damage has occurred are excluded since no reliable reconstruction can be performed due to severe plastic deformation in such regions.
Within the remaining regions, voids are identified via an object detection algorithm and reconstructed using a reverse erosion algorithm.
The fundamental principle of this approach is that the grain orientations vary smoothly within each individual grain, so void regions can be reliably constructed by propagating orientation information inwards from the surrounding undamaged material, one pixel layer per iteration, until the void is completely filled.
The GBEs are then calculated for all boundaries using the lattice matching method \cite{runnels2016analytical}.

The inputs to the ML model consist of reconstructed and repaired grain orientations and GBE (\cref{fig:ml_test}, rows 2-3).
To prevent gimbal lock, quaternions are used to parameterize SO(3), so that orientation is encoded in four continuous channels.
GBE is carried in a single additional channel, and all five channels are normalized.
The model is trained on mini-segments of \textit{post-mortem} microstructures, cropped with partial overlap.
Microstructures containing large voids are not considered.

For supervised learning, susceptibility maps are created from micrographs to serve as ground truth (\cref{fig:ml_test}, row 4).
These maps are generated using a convolved mask function
\begin{equation}\label{eq:groundtruth}
  D(x,y) = \left(M * \mathcal{G}_{\sigma}\right)\cdot\left(G * \mathcal{G}_{\epsilon}\right),
\end{equation}
where $M$ is the segmented void map, $G$ is the GBE field, $\mathcal{G}_{\sigma}$ and $\mathcal{G}_{\epsilon}$ are Gaussian mollifiers with standard deviations $\sigma$ and $\epsilon$ representing uncertainty in void location and boundary location, respectively, and the product is taken pointwise.
The results indicate the possible locations where voids can nucleate along the pristine boundaries, with the positional uncertainty reflected through mollification.
It is emphasized that such a construction limits the prediction to voids on grain boundaries.
Voids appearing entirely in the interior of a grain are completely eliminated.
This reduces the stochasticity of the dataset and reduces the risk of spurious influence by voids that nucleated far from the cut surface and were sectioned eccentrically.
Consequently, intragranular nucleation is not represented in the learned field.

For mapping these mini-microstructures to a void nucleation susceptibility field, a U-Net architecture is incorporated.
Further, the U-Net is enhanced with attention gates so that only the most salient encoder features contribute to the reconstruction of the void nucleation susceptibility field.
The model is evaluated using leave-one-out cross validation.
In each partition, all mini-microstructures originating from a single micrograph are assigned to the test set while the mini-microstructures from the remaining 56 micrographs constitute the training set, and repeating this procedure for every single micrograph yields 57 independent train-test splits.
This is done to ensure that the test data is always drawn from a completely different micrograph than the data on which the model is trained so the model learns to generalize across microstructures rather than within one.
Though this reduces the measured performance relative to a random split, it significantly enhances the reliability of the model.
Although the model is trained and tested on mini-microstructures, the learned mappings between microstructural features and void nucleation susceptibility fields are transferable to larger microstructural images. 

The void nucleation susceptibility fields predicted by the ML model are validated against the ground truth, obtained from actual experimental micrographs.
It should be noted that these predictions are made against unseen data.
The agreement between predictions and ground truth is quantified using the Pearson correlation coefficient (PCC) of the two images.
The PCC is used since the classic image metrics give misleadingly high scores.
In general, predicted void nucleation fields agree remarkably well with ground truth susceptibility fields (\cref{fig:ml_test}, row 5).
The predicted fields not only localize the void nucleated grain boundaries accurately but also predict potential void nucleation sites within the microstructure.

\section{Computational domain}\label{sec:domain}
\begin{figure}
    \centering
    \begin{subfigure}{1.02\textwidth}
        \centering
        \includegraphics[width=\textwidth]
{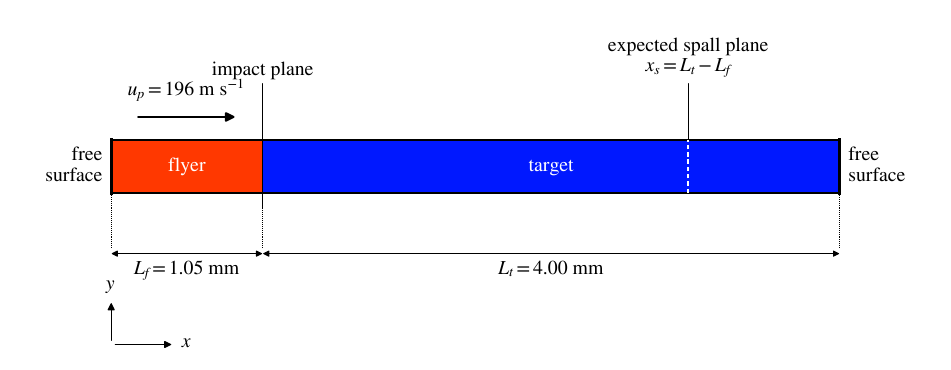}
        \caption{Geometry and initial velocity field. 
        The $1.05$~mm flyer occupying the left portion of the domain is assigned a uniform initial velocity of $\SI{196}{\meter\per\second}$ in the shock direction while the $4.00$~mm target is initially at rest. 
        The rear surfaces of both the flyer and target are free surfaces. 
        The expected spall plane lies one flyer thickness from the rear free surface of the target.}
        \label{fig:domain_vel}
    \end{subfigure}
    \\[1.5ex]
    \begin{subfigure}{0.49\textwidth}
        \centering
        \includegraphics[width=\textwidth]
{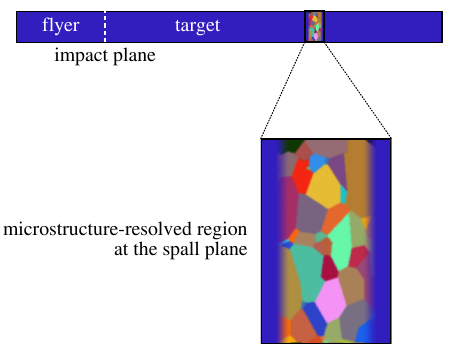}
        \caption{The polycrystalline microstructure is resolved within a thin band at the spall plane and transitions into a homogeneous material in the remainder of the domain. This polycrystalline microstructure comes from a pristine region of the experimental incipient spall micrograph.}
        \label{fig:domain_micro}
    \end{subfigure}
    \hfill
    \begin{subfigure}{0.49\textwidth}
        \centering
        \includegraphics[width=\textwidth]
{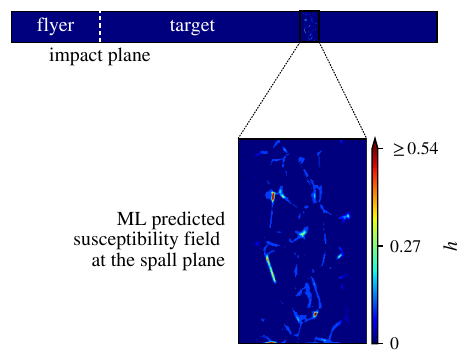}
        \caption{ML predicted nucleation susceptibility field $h$ for the microstructure resolved in the spall plane using grain orientations and GBE.
    The color scale is saturated at the 99th percentile of the nonzero values.
    Susceptibility of void nucleation is concentrated along grain boundaries and close to triple junctions.}
        \label{fig:domain_heatmap}
    \end{subfigure}
    \caption{Initial configuration of the computational domain. }
    \label{fig:domain}
\end{figure}

The computational domain uses the same flyer and target thicknesses as those used in the experiment \cite{jones2018effect}.
The computational domain is $\SI{5.05}{\milli\meter}$ long in the shock direction and $\SI{239}{\micro\meter}$ in the transverse direction, and is divided into a $\SI{1.05}{\milli\meter}$ flyer occupying the left portion of the domain and a $\SI{4}{\milli\meter}$ target occupying the remainder (\cref{fig:domain}). 
Both the flyer and the target are tantalum, corresponding to the experiment. 
To simulate the flyer impact on the target, the flyer region is initialized to a uniform velocity of $\SI{196}{\meter\per\second}$ in the shock direction.
The target is initially at rest.
The flyer and the target remain in contact and do not separate through the entirety of the simulation.
Because the flyer and target are represented within a single Lagrangian mesh, the interface transmits the impact without a contact algorithm or plasticity at the interface. 
Traction free boundary conditions are applied at both ends of the domain in the shock direction.
Therefore, the rear surfaces of the flyer and the target are both free surfaces. 
The free surface of the target provides the FSV record used for comparison with experiment.
Periodic boundary conditions are applied in the transverse direction.
Homogeneous Neumann conditions are applied to the damage field on all domain boundaries. 
Under these initial configurations and boundary conditions, the rarefaction fans originating from reflection of the shock waves from the two free surfaces interact at a depth of approximately one flyer thickness from the target free surface.
This is the plane where spallation occurs.

The transverse width of the domain is much smaller than the 19.1 mm diameter of the experimental target.
In plate impact experiments, the diameter of the targets is much larger than their thickness, so the material near the axis remains in a state of uniaxial strain \cite{meyers1994dynamic,antoun2003spall}.
In the experiment, the target is press-fitted into a tantalum ring of $64$~mm diameter, which prevents the release waves from the lateral edges from entering the target and perturbing the state of uniaxial strain \cite{jones2018effect}.
The FSV is measured at the center of the target and is therefore dependent on the thickness of the flyer and the target and not on the lateral dimension \cite{jones2018effect}.
However, a domain with the full diameter of the target would be computationally very expensive.
Therefore in the present approach, the thickness of the flyer and target is retained but the transverse width is set by the transverse width of experimental micrographs.
Further, the application of periodic boundary conditions on the top and bottom edges of the domain ensures that the domain is equivalent to a plate of infinite lateral extent, so the state of uniaxial strain is reproduced in the mean at any transverse width, while transverse heterogeneity is permitted at the scale of the resolved microstructure.

Because void nucleation is dependent on the local microstructural features, the nucleation criterion requires a resolved microstructure in the spall plane, where damage is expected to occur. 
Resolving the microstructure throughout the entire domain would require a fine mesh and would therefore be computationally very expensive.
Further, the wave propagation in the domain away from the spall region can be adequately represented by a homogeneous material.
Therefore, the microstructure is selectively resolved only within a band of finite thickness centered on the spall plane and the remainder of the domain is treated as homogeneous.
An abrupt change from the microstructure resolved region to the homogeneous region would introduce a mismatch at the edges of the band, which would partially reflect the shock and release waves.
Two measures are incorporated to avoid these reflections.
First, the orientation of the homogeneous region is chosen as an average of the orientations in the microstructure resolved region so that the longitudinal wave speed in both these regions closely match each other. 
Second, a transition zone of width $w_t$ is added on each side of the microstructure resolved region.
Let $a(x,y)$ denote either the orientation field or the void nucleation susceptibility field after the micrograph is placed, and let $a_m$ denote its value in the matrix.
The field used in the simulation is
\begin{equation}\label{eq:dissolve}
  \tilde{a}(x,y) = \bigl[1 - s(x)\bigr]\,\bigl(a * \mathcal{G}_{\varsigma(x)}\bigr)(x,y) + s(x)\,a_m ,
  \qquad \varsigma(x) = \varsigma_{\max}\, s(x),
\end{equation}
where $\mathcal{G}_{\varsigma}$ is a Gaussian mollifier with standard deviation $\varsigma$ and $s(x)$ is a weight function that increases smoothly from zero at the edge of the band to one at the end of the transition zone,
\begin{equation}\label{eq:smoothstep}
  s(x) = 3t^2 - 2t^3 ,
  \qquad
  t = \min\!\left[\max\!\left(\frac{|x - x_s| - w_b/2}{w_t},\,0\right),\,1\right].
\end{equation}
Here $x_s$ is the position of the spall plane, $w_b$ is the thickness of the band, $w_t$ is the width of the transition zone, and $\varsigma$ is the maximum width of the mollifier.
Inside the band $s = 0$ and the micrograph is unchanged, and beyond the transition zone $s = 1$ and the field equals its matrix value.
The matrix value of the void nucleation susceptibility field is zero.
Within this transition region, the orientation field is progressively smoothed with a Gaussian filter and blended into the homogeneous matrix orientation with a smooth weight function, so that the grains dissolve gradually into the matrix.

The micrographs embedded in the spall region are taken from the samples recovered from the plate impact experiments \cite{jones2018effect}.
These micrographs are processed with the help of a  microstructure reconstruction technique to reconstruct void regions.
Two considerations restrict the region of micrographs used in the spall plane of the domain.
First, regions containing large voids are not reliably reconstructed, since the material surrounding these voids has undergone severe plastic distortion. 
Second, when such a region is supplied into the ML model, the predicted void nucleation susceptibility field localizes at large voids without making predictions using microstructural features since voids are already present in inputs. 
Prescribing these fields in the simulation would suppress the interaction between local stresses and microstructural features that govern where voids should nucleate.
Therefore portions of micrographs away from the spall region, which are relatively pristine, with no visible voids, are chosen to be used in simulations.

\begin{figure}
    \centering
    \begin{subfigure}[t]{0.30\textwidth}
        \centering
        \includegraphics[width=\textwidth]
{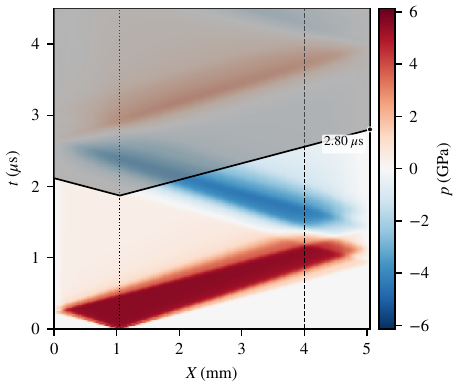}
        \includegraphics[width=\textwidth]
{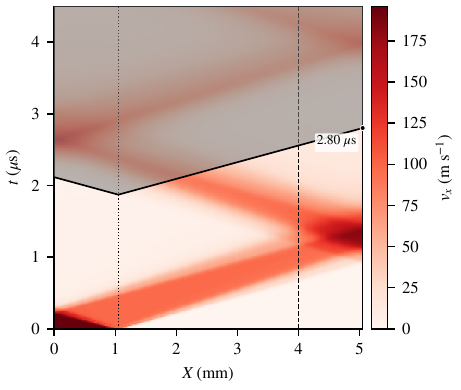}
        \caption{
        Response in the homogeneous reference domain.
        }
        \label{fig:xt_p_macro}
    \end{subfigure}
    \hfill
    \begin{subfigure}[t]{0.30\textwidth}
        \centering
        \includegraphics[width=\textwidth]
{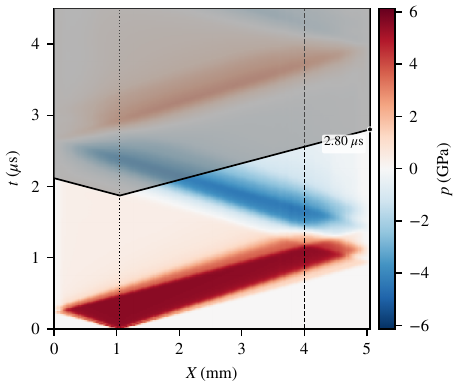}
        \includegraphics[width=\textwidth]
{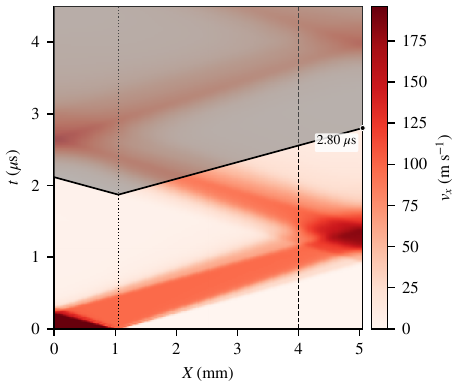}
        \caption{
        Response in the microstructure-resolved domain
        }
        \label{fig:xt_p_micro}
    \end{subfigure}
    \hfill
    \begin{subfigure}[t]{0.30\textwidth}
        \centering
        \includegraphics[width=\textwidth]
{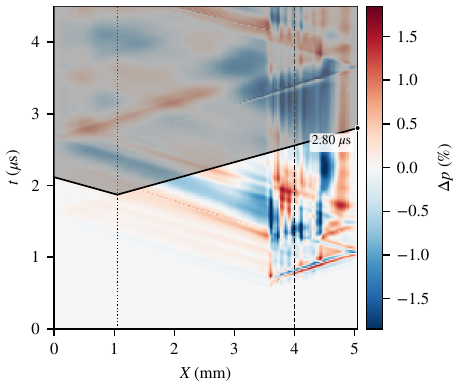}
        \includegraphics[width=\textwidth]
{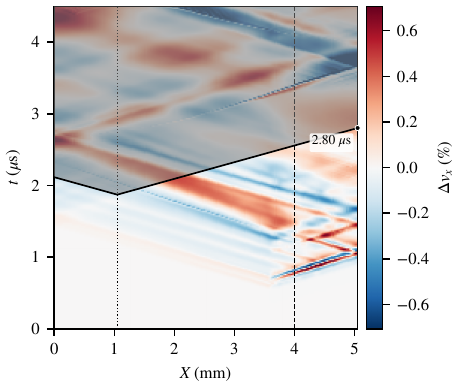}
        \caption{Difference between homogeneous and microstructure-resolved domains.}
        \label{fig:xt_dp}
    \end{subfigure}
    \caption{Comparison of $X-t$ diagrams for the pristine domain, microstructure-resolved domain, and relative difference between the two, demonstrating the effect of including microstructure awareness.
    The shaded region marks the part of the solution affected by the bonded interface between the flyer and the target. 
    The marker at \SI{2.8}{\micro\second} gives the time at which the forward characteristic reaches the rear free surface of the target without the bonded interface between the flyer and the target.
    }
    \label{fig:xt_p}
\end{figure}

To verify that resolving the microstructure only in the vicinity of the spall plane does not alter the bulk wave dynamics, a comparison is conducted between the $x$-$t$ diagram of the fully homogenized macroscale domain, which serves as the reference, and that of the hybrid domain in which the polycrystalline microstructure is resolved only at the spall plane (\cref{fig:xt_p}). 
Qualitative comparisons of the $x$-$t$ diagrams indicate a match in the trajectories of the elastic precursor, plastic shock front and reflection of shock waves from the free surface.
Quantitative comparisons give a clear picture of shock behavior within the microstructure resolved region.
Subtracting the homogenized fields from those of the resolved microstructure domain yields the difference in the  pressure and particle velocities. 
The largest differences in pressure is 0.6\% of the peak pressure and the largest difference in velocity is 1.5\% of the peak velocity.
These differences are on a scale several orders below the magnitude of the field values. 
The largest differences occur within the resolved microstructure bands first as the plastic shock front traverses the region and then again when release fans interact at these bands.
A weak scattered field propagates outwards from the microstructure resolved band along the characteristics and it is discernible throughout the rest of the domain but it has very low magnitudes as compared to the actual field values.
Overall, the magnitude of the fields in the resolved band closely matches that of the homogenized domain, and therefore the resolution of the microstructure restricted to the spall plane does not affect the bulk wave dynamics and helps reduce the computational cost. 

The $x$-$t$ diagram also shows a limitation of the bonded interface between the flyer and the target.
The release wave from the target free surface travels back through the target and reaches the flyer-target interface, and the tensile region extends beyond the interface into the flyer.
In the experiment, the flyer separates from the target once the interface is loaded in tension and the wave reflects at the interface.
In the simulation, however, the wave continues into the flyer and reflects at its rear free surface, which delays the reflected wave by $2L_f/c_L$, where $L_f$ is the flyer thickness and $c_L$ is the longitudinal sound speed.
The earliest arrival of the reflected wave follows the characteristic that leaves the interface at $t_d$.
For a separated flyer, this wave would reach the spall plane at $t_{sp} = t_d + (L_t - L_f)/c_L = \SI{2.56}{\micro\second}$ and the target free surface at $t^{*} = t_d + L_t/c_L = \SI{2.8}{\micro\second}$ where $L_t$ is the target thickness.
However, in the simulation, the same wave reaches the spall plane at $t_{sp} + 2L_f/c_L = \SI{3.05}{\micro\second}$ and the target free surface at $t^{*} + 2L_f/c_L = \SI{3.29}{\micro\second}$.
Therefore, the fields in the spall plane before $t_{sp}$ are unaffected by the bonded interface.

\section{Implementation}\label{sec:implementation}
The shock physics formulation, the dynamic crystal plasticity model, and the continuum damage model are implemented in Alamo \cite{runnels2025alamo}.
Alamo is based on the block-structured adaptive mesh refinement (BSAMR) library AMReX \cite{zhang2019amrex}.
The BSAMR strategy divides the entire domain into multiple levels of refinement, with each level being evolved independently.
Communication between levels is handled by averaging the fine solution onto the coarse level and by interpolating the coarse solution onto the boundaries of the fine patches \cite{runnels2021massively}.
Temporal subcycling is used to evolve each level at its own time step.
 
Selective refinement is used to make sure that there is adequate resolution only in the region of interest without incurring excessive computational cost.
In the present work, the region of interest is the spall plane, where the microstructure and voids must be resolved by the grid.
The remainder of the target and the flyer is homogeneous and carries just the shock wave and the release waves, which are captured on a coarse grid.
BSAMR is therefore used to refine the mesh only in the spall region where the microstructure is resolved, while the rest of the domain is coarse.
Since the grid moves with the material, this refined region remains fixed in the material throughout the simulation.

A nucleation event causes a void of radius $r_n$ to be inserted at the center of the nucleating cell.
The damage variable $d$ is set to 1 within every cell within $r_n$.
The gradient term of \cref{eq:drivingforce} forms a diffuse boundary of width $2\ell$ around the void region.
The radius $r_n$ in this work is set by the mesh and spans roughly 4 cells at the finest refinement level.
A void in tantalum is in the order of nanometers at the time of nucleation. 
Resolving such a small void would require a mesh about 2-3 orders finer than the present one.
Such fine resolutions would be computationally very expensive and therefore it is assumed in this work that nucleated voids have already grown from the nano length scale to the micron length scale.

\section{Results}\label{sec:Results}

\begin{figure}
    \centering
    \begin{subfigure}[t]{0.48\textwidth}
        \centering
        \includegraphics[width=\linewidth]
{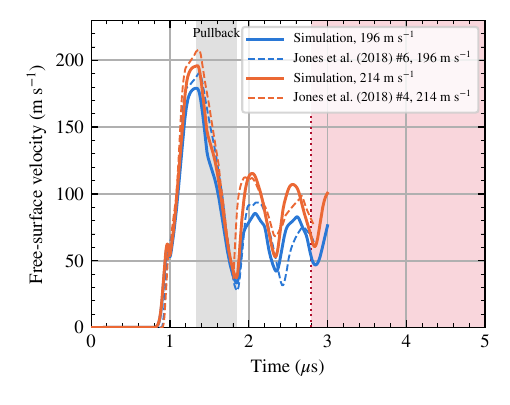}
        \caption{ Free-surface velocity over time}
        \label{fig:fsv_full}
    \end{subfigure}
    \hfill
    \begin{subfigure}[t]{0.48\textwidth}
        \centering
        \includegraphics[width=\linewidth]
{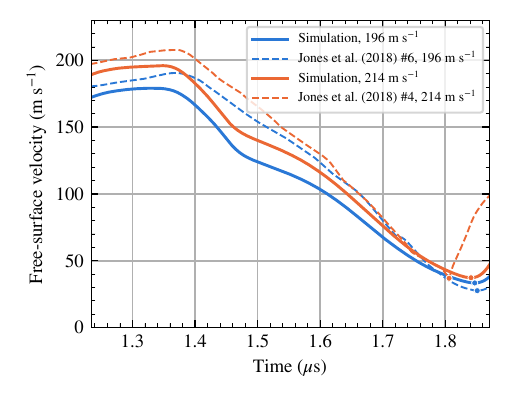}
        \caption{ Pullback region }
        \label{fig:fsv_pullback}
    \end{subfigure}
    \caption{
    FSV and pullback in comparison with the plate-impact experiments at $\SI{196}{\meter\per\second}$ and $\SI{214}{\meter\per\second}$ \cite{jones2018effect}.
    Simulations are solid lines and experiments are dashed lines.
    The simulated FSV agrees well with the $\SI{196}{\meter\per\second}$ experiment. 
    The HEL, pullback and the first reload are well captured.
    The gray band marks the pullback region shown in  \cref{fig:fsv_pullback}.
    The flyer and the target are bonded in this simulation but separate in reality at about $\SI{1.87}{\micro\second}$.
    Upon the separation, the resulting characteristic would reach the free surface of the target at $\SI{2.8}{\micro\second}$ at the earliest.
    Results after this time are affected by the bonded interface between the flyer and target and are shaded in red and not used for comparisons.
    Further, the limited accuracy of the coalescence model may also be contributing to the deviation between the experimental and simulated FSV. 
    Validation against the $\SI{214}{\meter\per\second}$ experiment shows a comparable level of agreement, and the same trends seen at $\SI{196}{\meter\per\second}$ are seen at $\SI{214}{\meter\per\second}$.
    }
    \label{fig:fsv_jones}
\end{figure}

\begin{figure}
    \centering
    \begin{subfigure}[t]{0.49\textwidth}
        \centering
        \includegraphics[width=\linewidth]
{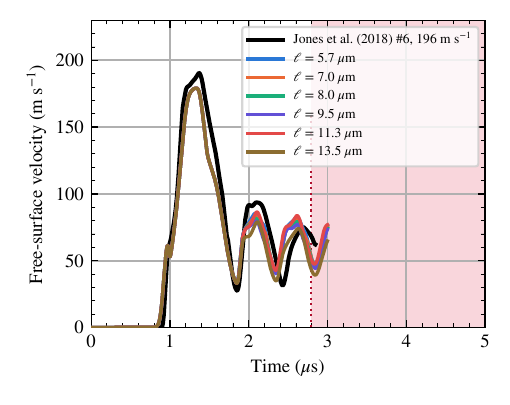}
        \caption{Regularization length $\ell$ (seed 1)}
        \label{fig:sens_ell}
    \end{subfigure}
    \hfill
    \begin{subfigure}[t]{0.49\textwidth}
        \centering
        \includegraphics[width=\linewidth]
{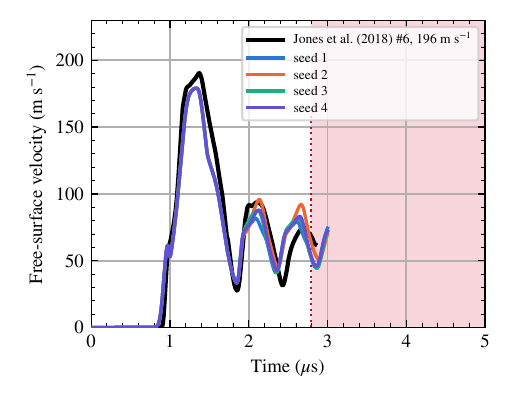}
        \caption{Random seed of the nucleation law ($\ell = \SI{8}{\micro\meter}$)}
        \label{fig:sens_seed}
    \end{subfigure}
    \caption{
    FSV sensitivity to the regularization length and random seeding at $\SI{196}{\meter\per\second}$.
    All simulations use a void radius of $\SI{8}{\micro\meter}$ with the regularization length varying by a factor of 2.8 (\cref{fig:sens_ell}).
    The pullback minimum and the spall strength do not vary with $\ell$ and seeds.
    There is a small variation in the first reload peak and the second reload peak with $\ell$ and seeds with the largest variation being observed for one of the random seeds (\cref{fig:sens_seed}).
    Overall, the variation in the FSV is minimal and the FSV is consistent for different values of $\ell$ and different seeds. }   
    \label{fig:fsv_sensitivity}
\end{figure}

\subsection{Comparison of free surface velocity and pullback with experimental results}
Since the FSV is a primary diagnostic available during plate impact experiments, the coupled continuum damage model and the crystal plasticity model are first validated against experimental FSV.
The FSV obtained from simulations is validated against experimental measurements from plate-impact tests on tantalum at an impact velocity of $\SI{196}{\meter\per\second}$ \cite{jones2018effect} (\cref{fig:fsv_jones}). 
In this experiment, a 1.05~mm flyer impacts a 4~mm target to produce a peak stress of 5.8~GPa and $\SI{6e4}{\per\second}$ strain rates \cite{jones2018effect}.
The simulated and measured FSV are aligned at the foot of the elastic precursor. 
It is emphasized that different parts of the FSV validate different components of the framework. 
From impact to the pullback minimum, the FSV is governed by the crystal plasticity model.
Beyond the pullback minimum, the FSV is governed by the damage model.

At $\SI{196}{\meter\per\second}$ the elastic precursor, shock arrival time and the time of peak of the FSV are captured within 3\% of the experimental values.
This indicates that the wave speeds within the flyer and the target and the timing of the tensile pulse are accurately captured by the model.
The peak velocity is 6\% lower than the experimental measurements and the release from the peak lies below the experimental measurements by similar fractions.
These early stage deviations can be attributed to the crystal plasticity model and the equation of state since no void nucleates in the domain at these times.
The pullback magnitude is reproduced within 11\% and the pullback minimum lies $\SI{6}{\meter\per\second}$ above experimental measurements, which is the deficit observed even in the peak velocity.
The time of the pullback minimum is reproduced within $\SI{4}{\nano\second}$.
Further, $\sigma_c$, $t_s$, $\mathbf{B}$ and $v_0t_0$ of the nucleation law are not experimental measurements.
These parameters are calibrated to match experimental $\SI{196}{\meter\per\second}$ FSV measurements.
Therefore, the FSV measurements on the $\SI{196}{\meter\per\second}$ case are not completely independent predictions but calibrated results.

Following the pullback minimum, the FSV captures the recompaction signal and subsequent reverberation of waves between the spall plane and the free surface of the target.
The first reload peak is 9\% lower than the experimental measurements and arrives within $\SI{25}{\nano\second}$ of the measured time.
The  modest deviation of amplitude and period from experimental measurements can be attributed to two factors.
First, the flyer and target remain unseparated throughout the simulation whereas they separate in the actual experiment.
This delays rarefaction waves from reaching the rear free surface of the target as compared to the expected time of $\SI{2.8}{\micro\second}$, in turn affecting the amplitude and period of the ringing.
Therefore the region of the FSV beyond  $\SI{2.8}{\micro\second}$ is not considered for comparisons.
Second, the damage model may not be accurately capturing the coalescence behavior of voids in the spall plane. 

To test whether the model calibrated on $\SI{196}{\meter\per\second}$ FSV transfers to other loading conditions, the FSV is also compared with the plate impact test at a flyer velocity of $\SI{214}{\meter\per\second}$.
All model parameters are retained from the $\SI{196}{\meter\per\second}$ case, while only changing the flyer velocity.
The same ML susceptibility field is used for this validation as for $\SI{196}{\meter\per\second}$ simulations.
This is based on the assumption that the microstructural features leading to void nucleation do not change with flyer velocity, so that only the stress term in the barrier of \cref{eq:nucleation} responds to higher flyer velocity.
The peak velocity is again 6\% low and the magnitude of the pullback is 7\% low, with the pullback minimum arriving $\SI{35}{\nano\second}$ later than measured values.
Further, the first reload is $\SI{88}{\nano\second}$ later than experimental measurements.
The calibration at $\SI{196}{\meter\per\second}$ therefore transfers to a  similar loading condition.
It is further noted that this does not remove the dependence of the $\SI{196}{\meter\per\second}$ case on the calibration of $\sigma_s$.

The damage model incorporates a regularization length and the nucleation criterion places voids at different sites depending on the local seed.
Both of these impact the damage profile in the spall plane.
Their effect on the FSV is studied with simulations at a flyer velocity of $\SI{196}{\meter\per\second}$.
The change in $\ell$ has very little effect on the FSV.
For $\ell$ between $\SI{5.7}{\micro\meter}$ and $\SI{13.5}{\micro\meter}$, the pullback minimum and its time are unchanged to within $\SI{0.3}{\meter\per\second}$ and $\SI{2}{\nano\second}$.
The first reload varies over a spread of 8\% of the measured value (\cref{fig:sens_ell}).
Further, the change in seeds has similar effects on the FSV.
Across the four seeds at $\ell = \SI{8}{\micro\meter}$, the pullback minimum changes by less than $\SI{0.3}{\meter\per\second}$ and its time by less than $\SI{3}{\nano\second}$ with little variations in the reload peak (\cref{fig:sens_seed}).
Overall, the FSVs have minimal variations with $\ell$ and seeds and this is the expected behavior.

It is noted that continuum damage models commonly represent the spall region with the help of a homogenized porosity field, and such models are generally calibrated to reproduce the FSV \cite{czarnota2006modelling,wilkerson2014dynamic}.
However, nearly identical FSV profiles can arise from  different damage profiles within the spall plane \cite{jones2018effect}.
Agreement in the FSV is therefore necessary but not sufficient to validate a spall model, as it does not indicate where voids nucleate or how damage is distributed in the microstructure.
Accordingly, the present framework is assessed not only through the FSV but also through the locations at which voids nucleate in the resolved microstructure (\cref{sec:results_overlay}).
\begin{figure}
  \centering
  \begin{subfigure}{0.5\linewidth}
    \centering
    \includegraphics[width=0.9\linewidth]
{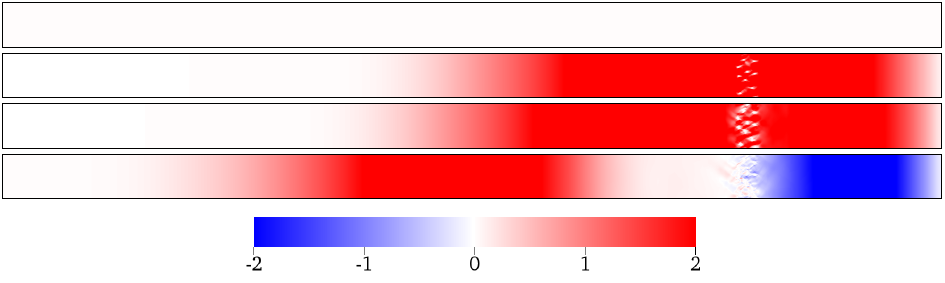}
    \caption{Axial stress $\sigma_{xx}$ (\si{\giga\pascal}).}
    \label{fig:macro_sxx}
  \end{subfigure}%
  \begin{subfigure}{0.5\linewidth}
    \centering
    \includegraphics[width=0.9\linewidth]
{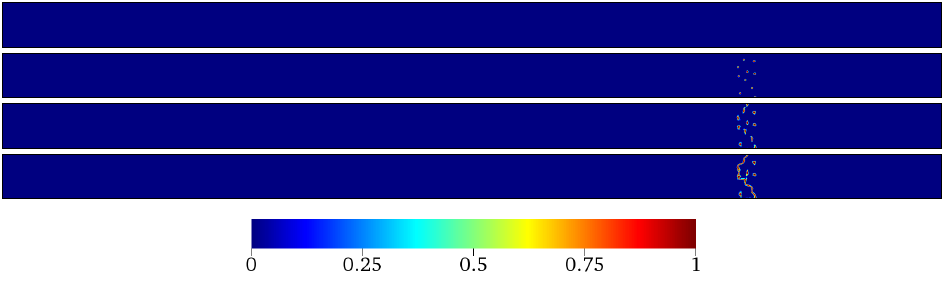}
    \caption{Damage $d$.}
    \label{fig:macro_d}
  \end{subfigure}
  \centering
  \begin{subfigure}{0.5\linewidth}
    \vspace{0.5cm}
    \centering
    \includegraphics[width=0.9\linewidth]
{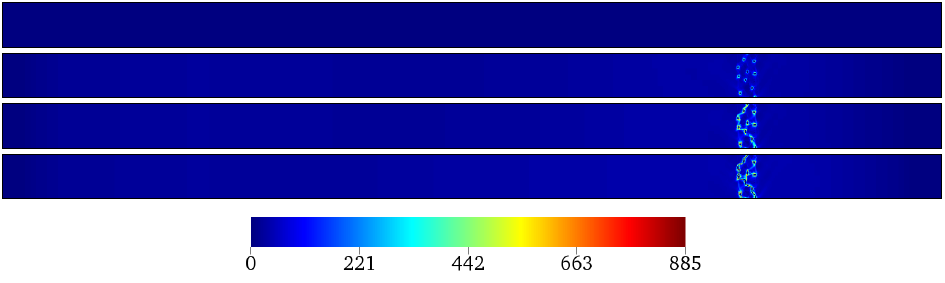}
    \caption{Accumulated plastic work density $W_p$ (\si{\mega\joule\per\cubic\meter}).}
    \label{fig:macro_wp}
  \end{subfigure}

  \caption{Full-domain fields of the Ta-on-Ta plate-impact simulation at
    $t = \SI{0}{\micro\second}$, \SI{1.59}{\micro\second},
    \SI{1.7}{\micro\second} and \SI{1.9}{\micro\second} (top to bottom in each panel). 
    The flyer occupies $0 \le x \le \SI{1.05}{\milli\meter}$ and
    the target occupies $\SI{1.05}{\milli\meter} < x \le \SI{5.05}{\milli\meter}$.
    Tension develops across the target and is sufficient to cause void nucleation by \SI{1.59}{\micro\second}. 
    Voids appear at the spall plane at $x \approx \SI{1.05}{\milli\meter}$ from the target free surface at \SI{1.59}{\micro\second}, and a continuous damage band spans the transverse width by \SI{1.9}{\micro\second} due to growth and coalescence mediated by plasticity.
    Plastic work concentrates around the voids and near ligaments at the damage band.}
  \label{fig:macro_filmstrip}
\end{figure}

\subsection{Void nucleation and damage growth at the spall plane}
While the FSV validates the overall material response, it is an indirect measure of damage and does not measure where voids nucleate in the spall plane and how they grow and coalesce.
To understand this, the axial stress, damage profile and the plastic work density are first examined at the macroscale over the entire domain (\cref{fig:macro_filmstrip}).
At $t=0$, the entire domain is stress free and undamaged.
Following the impact, stress builds up in the flyer and the target.
Once the shock waves reflect from the free surface as rarefaction waves, they interact at a distance roughly equal to the flyer thickness and produce a state of tension (\cref{fig:macro_sxx}).
At  $\SI{1.59}{\micro\second}$, this tension results in the nucleation of voids in the spall plane (\cref{fig:macro_d}).
Further, plasticity is observed around nucleated voids, which eventually drives growth (\cref{fig:macro_wp}).
After nucleation of voids, plasticity continues to drive void growth and coalescence with plastic work density increasing around voids at $\SI{1.7}{\micro\second}$.
The voids coalesce into a damage band spanning almost the entire transverse width by  $\SI{1.9}{\micro\second}$.
It is noted that the damage and plastic work remain localized to the spall region, while the rest of the domain remains undamaged for the entire duration of the simulation.

\begin{figure}
  \centering
  \begin{subfigure}{0.75\textwidth}
    \centering
    \includegraphics[width=\linewidth]
{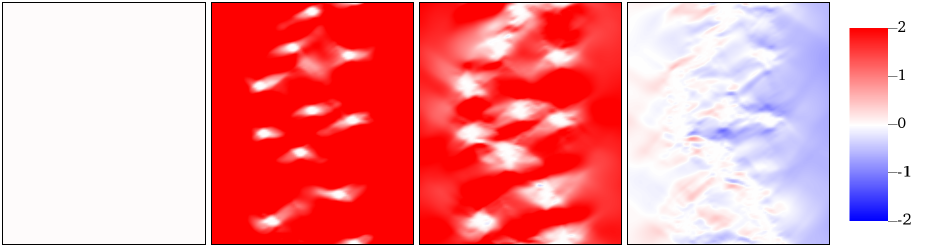}
    \caption{Axial stress $\sigma_{xx}$ (\si{\giga\pascal}).}
    \label{fig:spall_sxx}
  \end{subfigure}

  \vspace{6pt}
  \begin{subfigure}{0.75\textwidth}
    \centering
    \includegraphics[width=\linewidth]
{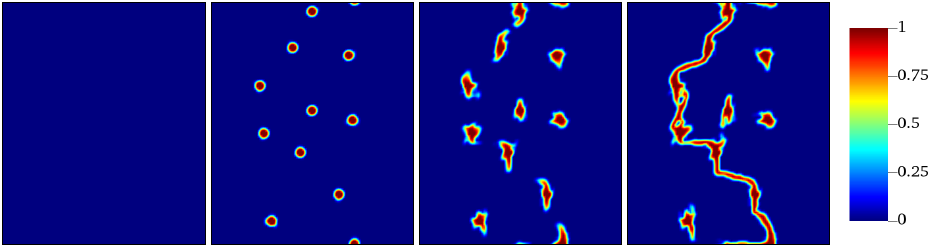}
    \caption{damage $d$.}
    \label{fig:spall_d}
  \end{subfigure}

  \vspace{6pt}
  \begin{subfigure}{0.75\textwidth}
    \centering
    \includegraphics[width=\linewidth]
{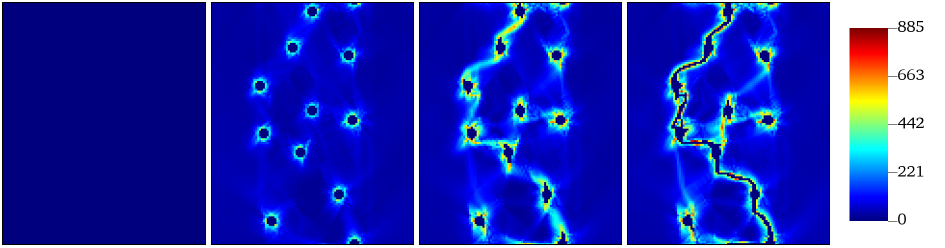}
    \caption{Accumulated plastic work density $W_p$ (\si{\mega\joule\per\cubic\meter}).}
    \label{fig:spall_wp}
  \end{subfigure}

  \caption{Spall region $\SI{2.89}{\milli\meter} \le x \le \SI{3.01}{\milli\meter}$  of \cref{fig:macro_filmstrip} over the full transverse width at $t = \SI{0}{\micro\second}$, \SI{1.59}{\micro\second}, \SI{1.7}{\micro\second} and \SI{1.9}{\micro\second} (left to right in each panel). 
  As tension builds up, voids nucleate at \SI{1.59}{\micro\second} to relieve the built up tension. (\cref{fig:spall_sxx}).
  Isolated voids formed at \SI{1.59}{\micro\second} grow and coalesce into a continuous damage band by \SI{1.9}{\micro\second} (\cref{fig:spall_d}). 
  Plastic work concentrates at the void rims and in the ligaments between voids and is responsible for growth and coalescence (\cref{fig:spall_wp}).
  }
  \label{fig:spall_filmstrip}
\end{figure}

Since the damage is localized in a narrow spall region, the same fields are examined in a magnified view of the spall region, in which the microstructure is resolved (\cref{fig:spall_filmstrip}). 
At $\SI{1.59}{\micro\second}$, the spall region is under tension and the nucleated voids are isolated from each other (\cref{fig:spall_d}).
Each nucleated void is surrounded by a thin diffuse interface.
The tensile stress is relieved locally around each void but there is still some stress in the ligament.
During the initial nucleation phase, the plastic work density is highest at void rims (\cref{fig:spall_wp}). 
Interestingly, while no voids nucleate between $\SI{1.59}{\micro\second}$ and $\SI{1.7}{\micro\second}$, the plastic work increases significantly in the ligament connecting neighboring voids.
This indicates that plasticity localizes in the ligaments between the voids before the voids link with each other.
By $\SI{1.9}{\micro\second}$, the voids have grown and coalesced along these ligaments into a continuous damage band.

It should be noted that this damage band observed from coalescence has voids around it that do not coalesce, a phenomenon observed in experimental micrographs \cite{jones2018effect}.
Further, the tension in the spall region relaxes to almost zero once voids coalesce and the plastic work density is highest around voids at this instant.
Overall, the model captures the sequence of void nucleation, growth and coalescence that is observed during spallation of ductile materials.

The transverse width of the computational domain is $\SI{239}{\micro\meter}$ and is set by the width of the pristine part of the recovered micrograph.
The width of the strip cannot be increased further since it would then enter the non-pristine regions of the micrographs that carry large voids.
If these void regions are not reliably reconstructed due to severe plastic deformation around them, they could potentially pollute machine learned susceptibility fields.
Since the transverse width of the computational domain is a very small part of the 19.1 mm diameter target used in the experiment, sufficient damage must develop in the domain to accurately recover the FSV.
To nucleate a sufficient number of voids, the parameter $\mathbf{B}$, which controls the barrier height, is modified along with the nucleation rate scale $v_0t_0$.
Further the parameter $r_{space}$ acts as a controlling parameter to limit the number of nucleation events.
Hence, the number of voids and therefore the extent of coalescence is a consequence of $\mathbf{B}$, $v_0t_0$ and $r_{space}.$

\begin{figure}
  \centering
  \begin{subfigure}[t]{0.4\textwidth}
    \centering
    \includegraphics[width=\linewidth]
{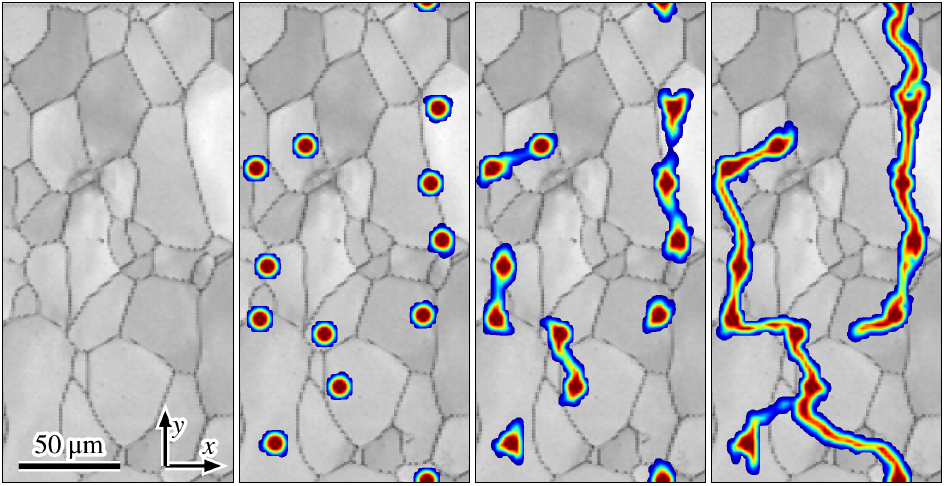}
  \end{subfigure}

  \vspace{4pt}
  \begin{subfigure}[t]{0.4\textwidth}
    \centering
    \includegraphics[width=\linewidth]
{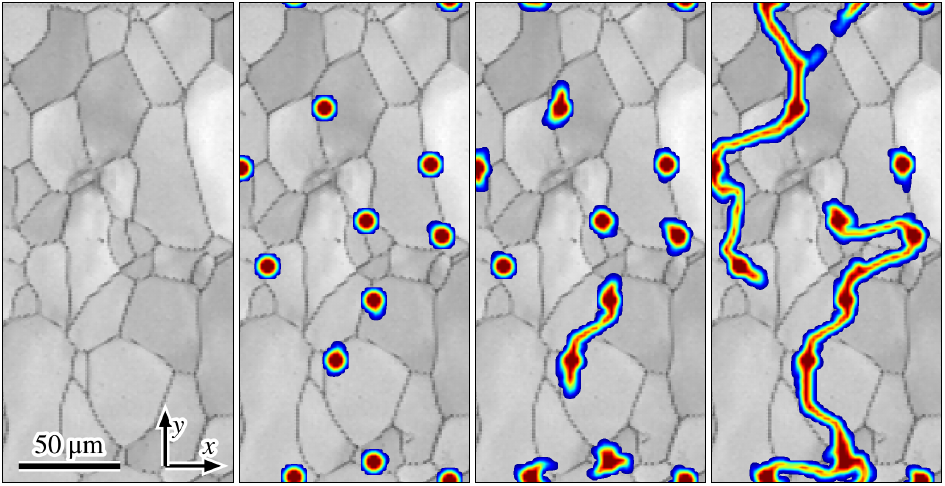}
  \end{subfigure}

  \vspace{4pt}
  \begin{subfigure}[t]{0.4\textwidth}
    \centering
    \includegraphics[width=\linewidth]
{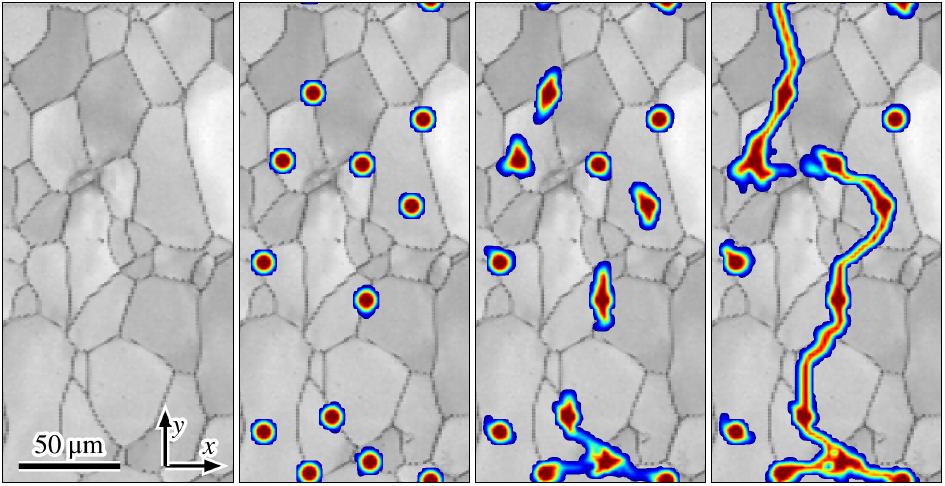}
  \end{subfigure}

  \vspace{4pt}
  \begin{subfigure}[t]{0.4\textwidth}
    \centering
    \includegraphics[width=\linewidth]
{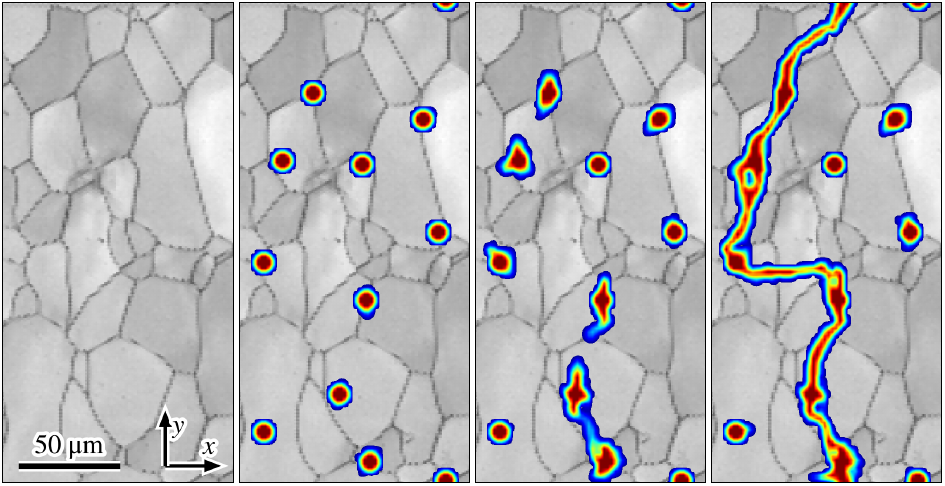}
  \end{subfigure}
  \caption{Simulated damage field overlaid on the resolved microstructure of the spall region for four different simulations. 
  Each row indicates damage profiles for one simulation.
  The four simulations share the same embedded micrograph, susceptibility field and loading conditions but differ in the random seed of the nucleation law.
  The first column indicates the micrograph embedded in the spall region of the domain, approximately $1.05$ mm from the free surface of the target and is identical in all rows.
  The second column indicates void nucleation at $t = \SI{1.61}{\micro\second}$.
  In each simulation, voids nucleate predominantly along grain boundaries with only one intragranular nucleation.
  The intragranular nucleation occurs in the same grain but at different locations within the grain in all simulations.
  In each simulation, some intergranular voids nucleate at the same location or along the same boundary irrespective of the initial seed, indicating that specific boundaries preferentially nucleate voids over other boundaries.
  The third column shows void growth at $t = \SI{1.7}{\micro\second}$.
  Each void grows at a different rate, highlighting the impact of plasticity on void growth.
  The fourth column shows void coalescence occurring at $t = \SI{1.9}{\micro\second}$.
  Damage bands preferentially tend to grow along specific grain boundaries and sometimes cut through grains. 
  In all simulations, values $d < 0.02$ are hidden for better visualization.}
  \label{fig:overlay}
\end{figure}

\subsection{Void nucleation sites and damage growth in the resolved microstructure}\label{sec:results_overlay}

To examine the effect of stochasticity on where voids nucleate in the microstructure and whether nucleation sites are repeatable across simulations, four representative configurations are compared.
These simulations are run with the same embedded microstructure, the same void nucleation susceptibility field, and the same loading condition, differing only in the random seed used to generate nucleations.
The damage field in each simulation is overlaid on the micrograph resolved in the spall plane at different stages of damage progression.
It is emphasized that the embedded micrograph is pristine and therefore the results indicate where voids nucleate and how damage progresses in a pristine microstructure.

By $\SI{1.61}{\micro\second}$, voids nucleate in all simulations, with majority of the voids nucleating along grain boundaries (\cref{fig:overlay}).
Only one void nucleates inside of a grain, but it is interesting to note that the intragranular void always nucleates in the same grain but at different locations within the grain (\cref{fig:overlay}).
Further, specific void nucleation sites at triple junctions remain common in all simulations irrespective of the initial random seed.
Also, specific grain boundaries tend to nucleate voids in all simulations, though the actual location of nucleation along the boundary varies for each simulation.
This clearly indicates that some boundaries preferentially nucleate voids over other boundaries, consistent with molecular dynamic simulations \cite{fensin2014some}.

By $\SI{1.7}{\micro\second}$,  while no new voids nucleate, the voids in the domain indicate growth (\cref{fig:overlay}).
Each void grows at a different rate indicating the impact of plastic work density on void growth.
Further voids with neighbors close to them preferentially grow along the ligament between them.
By $\SI{1.9}{\micro\second}$, voids have coalesced into a continuous damage band in all four simulations.
These bands tend to follow specific grain boundaries but sometimes cut across grain interiors.
It should be noted that the damage model itself carries no information about the location of grain boundaries.
The path of the band is set by the location of the nucleation sites and by the stress and plasticity surrounding voids.

All these different damage profiles produce similar FSVs agreeing with experimental observations \cite{jones2018effect}.
Additionally, nucleation of voids predominantly at grain boundaries is consistent with intergranular void nucleation observed in recovered samples.
It would be interesting to qualitatively compare damage profiles with post-mortem micrographs but that remains a subject for future work.

\subsection{Microstructure effects on damage and FSV}\label{sec:micro_effects}

To isolate microstructural effects on void nucleation, simulations are performed by replacing the microstructure in the spall plane with synthetic microstructures while the loading conditions, domain and other model parameters are held fixed.
The effect of each microstructure is then assessed via the FSV.

\begin{figure}
  \centering
  \captionsetup[subfigure]{justification=centering,singlelinecheck=false,font=scriptsize,skip=1pt}
  \begin{subfigure}[b]{0.53\linewidth}
    \newlength{\panelw}\setlength{\panelw}{0.27\linewidth}
    \newlength{\panelgap}\setlength{\panelgap}{0.012\linewidth}
    \newlength{\labelw}\setlength{\labelw}{0.15\linewidth}
    \newlength{\labeloff}\setlength{\labeloff}{\dimexpr0.5\panelw-1.1em\relax}
    \scriptsize
    \makebox[\labelw]{}\hspace{\panelgap}%
    \makebox[\panelw]{\textbf{Orientation}}\hspace{\panelgap}%
    \makebox[\panelw]{\textbf{GBE}}\hspace{\panelgap}%
    \makebox[\panelw]{\textbf{Prediction}}%
    \par\vspace{3pt}\noindent
    \begin{minipage}[t]{\labelw}\vspace{0pt}\vspace{\labeloff}
      \raggedleft\textbf{Single\\crystal}
    \end{minipage}\hspace{\panelgap}%
    \begin{minipage}[t]{\panelw}\vspace{0pt}
      \includegraphics[width=\linewidth]
{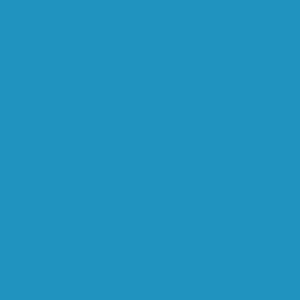}
    \end{minipage}\hspace{\panelgap}%
    \begin{minipage}[t]{\panelw}\vspace{0pt}
      \includegraphics[width=\linewidth]
{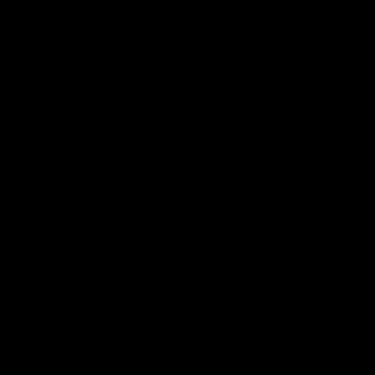}
    \end{minipage}\hspace{\panelgap}%
    \begin{minipage}[t]{\panelw}\vspace{0pt}
      \includegraphics[width=\linewidth]
{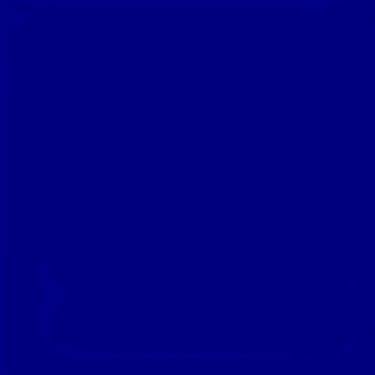}
    \end{minipage}%
    \par\vspace{3pt}\noindent
    \begin{minipage}[t]{\labelw}\vspace{0pt}\vspace{\labeloff}
      \raggedleft\textbf{Voronoi\\polycrystal}
    \end{minipage}\hspace{\panelgap}%
    \begin{minipage}[t]{\panelw}\vspace{0pt}
      \includegraphics[width=\linewidth]
{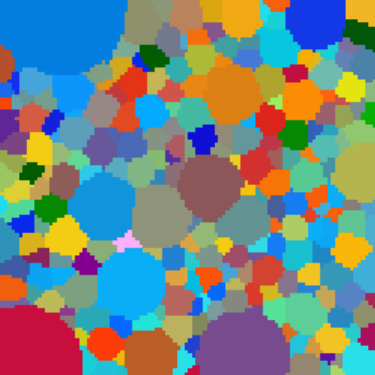}
    \end{minipage}\hspace{\panelgap}%
    \begin{minipage}[t]{\panelw}\vspace{0pt}
      \includegraphics[width=\linewidth]
{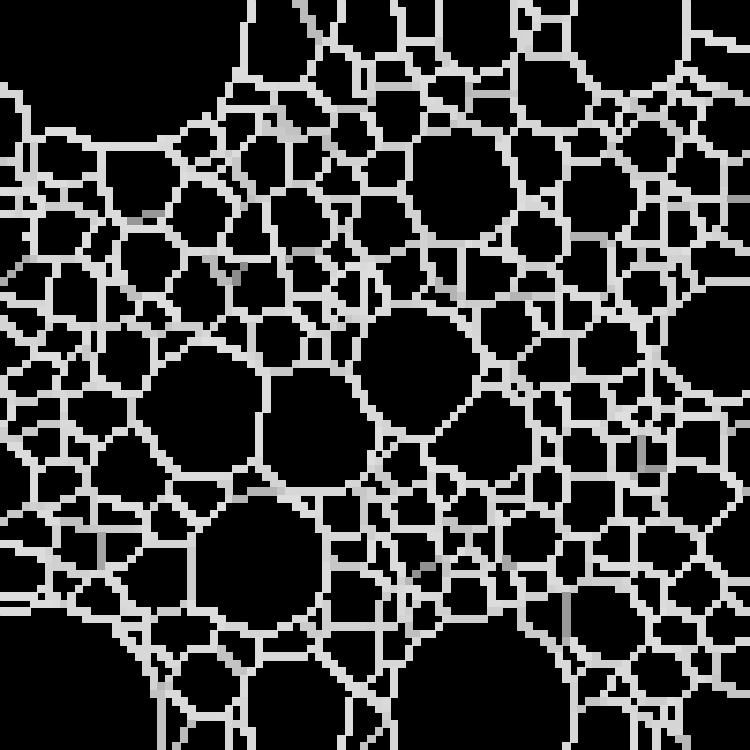}
    \end{minipage}\hspace{\panelgap}%
    \begin{minipage}[t]{\panelw}\vspace{0pt}
      \includegraphics[width=\linewidth]
{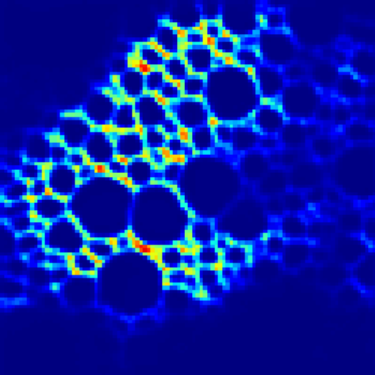}
    \end{minipage}%
    \par\vspace{24pt}
    \caption{Synthetic microstructures}\label{fig:microstructure_grid}
    \label{fig:sc_orient}\label{fig:sc_gbe}\label{fig:sc_heatmap}
    \label{fig:vor_orient}\label{fig:vor_gbe}\label{fig:vor_heatmap}
  \end{subfigure}\hfill%
  \begin{subfigure}[b]{0.45\linewidth}
    \includegraphics[width=\linewidth]
{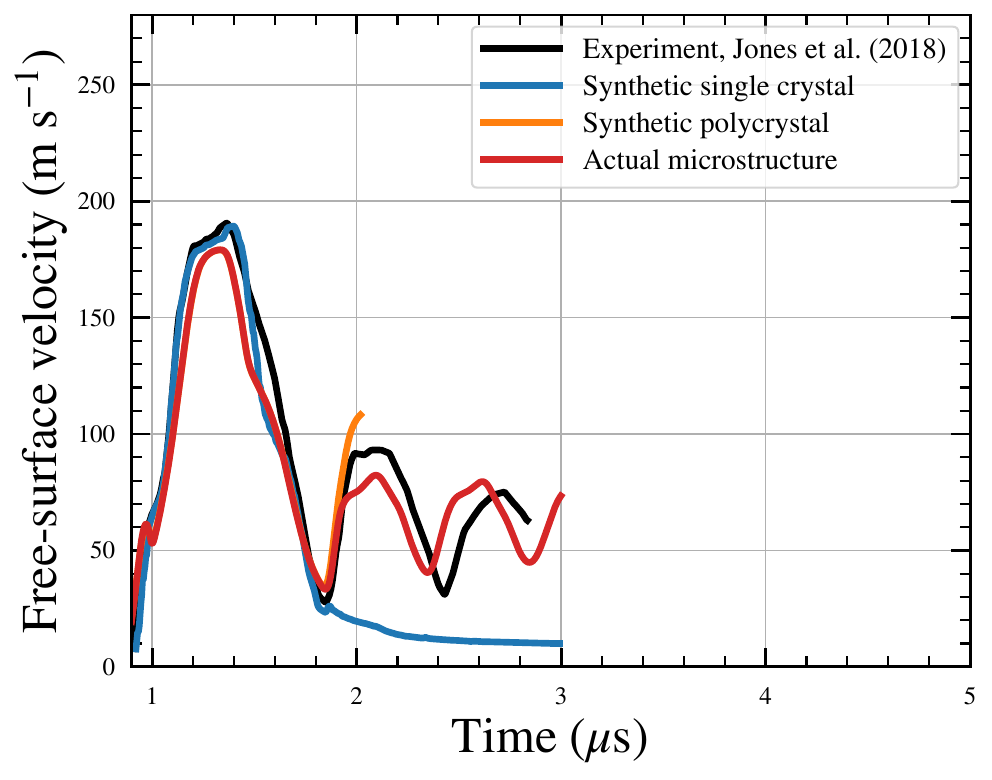}
    \caption{Relative FSV responses}\label{fig:micro_fsv}
  \end{subfigure}
  \caption{Synthetic microstructures embedded in the spall region.
  (\subref{fig:microstructure_grid}) Grain orientation (left), GBE (middle), and void nucleation susceptibility fields predicted by the ML model (right) for the single crystal (top) and the Voronoi polycrystal (bottom).
  The single crystal has no grain boundaries, so its GBE is zero everywhere.
  The Voronoi polycrystal contains 240 grains of mean diameter 5~\textmu m, with orientations sampled from the orientation distribution of the experimental micrographs \cite{jones2018effect}.
  Its GBE is computed by the lattice matching method \cite{runnels2016analytical}.
  (\subref{fig:micro_fsv}) FSV for the single crystal and the Voronoi polycrystal at an impact velocity of $\SI{196}{\meter\per\second}$.
  In the single crystal, no voids nucleate and no spall signal appears.
  The Voronoi polycrystal reproduces the pullback signal, with substantially higher ringing amplitude than the experimental micrograph.
  The Voronoi simulation is terminated at $\SI{2.02}{\micro\second}$, after the onset of the ringing.
  The actual microstructure on the other hand reproduces the FSV accurately as compared to the synthetic microstructures.}
  \label{fig:micro_inputs}
\end{figure}

The first synthetic microstructure is a single crystal embedded within the spall region.
Since the single crystal contains no grain boundaries, the ML model predicts no susceptibility field across the entire crystal.
Further, the stress term in the nucleation criterion remains active, but it fails to nucleate voids in the single crystal.
Accordingly, the simulated FSV produces no spall signal (\cref{fig:micro_inputs}).
Following the release from the peak, the signal decreases past the pullback minimum and no reverberation between the spall plane and the target free surface is observed.
The absence of a reload in the FSV indicates higher spall strength of the single crystal relative to the polycrystal \cite{remington2018spall}.
However, intragranular void nucleation criterion is only represented through the stress term of the void nucleation criterion.
Therefore, whether these single crystal simulations accurately represent spall behavior in single crystals is not well established.

The second synthetic microstructure is a polycrystal generated by a Voronoi tessellation, with 240 grains of mean diameter $\SI{5}{\micro\meter}$, much smaller than the mean grain size of approximately $\SI{30}{\micro\meter}$ in the experimental material \cite{jones2018effect}.

Grain orientations are sampled from the orientation distribution measured in the experimental micrographs \cite{jones2018effect}.
Therefore the synthetic microstructure and the experimental microstructure share the same grain orientation distribution but differ in the grain boundary network. 
The GBE is computed for all boundaries using the lattice matching method \cite{runnels2016analytical}. 
The orientation and GBE fields are fed into the trained ML model to obtain the void nucleation susceptibility field.
Since the average grain size is only $\SI{5}{\micro\meter}$, the void radius is reduced to $\SI{1.6}{\micro\meter}$ from $\SI{5}{\micro\meter}$ used in simulations with experimental microstructures.
All other parameters remain the same.
The synthetic polycrystal contains a higher density of grain boundaries than the experimental micrograph.
Accordingly, the void nucleation susceptibility is nonzero over a larger fraction of the spall region, and a larger number of voids nucleate.
The simulated FSV reproduces the pullback signal with a small difference as compared with the experiment.
The small difference in magnitude between the two can be attributed to the difference in their microstructures. 
The first reload overshoots the experimental measurements even though nucleated voids are smaller compared to simulations using experimental microstructures in the spall plane.
This indicates that many small voids nucleate close to each other and coalesce rapidly.
The higher void density and therefore faster coalescence, and the spall region reaches complete separation much faster as compared to experimental microstructures.
A fully separated spall plane acts as a free surface for the waves trapped between the spall plane and the target free surface, so that a larger fraction of each wave is reflected and the ringing amplitude increases.
These results clearly indicate that for the cases examined, the pullback signal is insensitive to the microstructure embedded in the spall region, whereas the late time ringing depends strongly on the density of nucleation sites and the resulting extent of coalescence.
This is consistent with the observation that nearly identical FSV profiles up to the pullback minimum can arise from markedly different distributions of damage within the spall region \cite{jones2018effect}.

\section{Conclusion}\label{sec:Conclusions}
A continuum damage framework for spallation in BCC tantalum under plate impact is presented in this work.
The framework couples a Lagrangian shock physics formulation, a dislocation density based crystal plasticity model and a continuum damage model for void growth.
Void nucleation is governed by a thermally activated rate law in which the barrier is the sum of a microstructural contribution and a mechanical contribution.
The microstructural contribution is obtained from an ML model that predicts a void nucleation susceptibility field from grain orientation and GBE.
The mechanical contribution depends on the maximum non negative principal Cauchy stress.
Voids therefore nucleate where a susceptible grain boundary and a sufficient maximum non negative principal stress occur together.
To reduce computational cost, the microstructure is resolved only within a band containing the spall plane, and the remainder of the domain is treated as a homogeneous material.

The $x$-$t$ diagrams of the domain with the resolved microstructure and of the fully homogenized domain show matching trajectories of the elastic precursor, the plastic wave, and the reflected rarefaction waves.
The largest differences between the two domains are 0.6\% of the peak pressure and 1.5\% of the peak particle velocity, and they are predominantly confined to the resolved microstructure band as the shock waves and release waves traverse it.
Restricting the resolved microstructure only to the spall plane therefore significantly reduces the computational cost without altering the bulk wave dynamics.

The simulated FSV at an impact velocity of $\SI{196}{\meter\per\second}$ is compared with plate-impact measurements on tantalum.
The elastic precursor, the shock arrival and the time of the peak are captured within 3\% of experimental measurements, which indicates that the wave speeds in the target and flyer are accurately represented.
The peak velocity is 6\% lower than experimental values and the release from the peak lies below experimental measurements by a similar fraction.
Since these deviations arise before voids nucleate, they are attributed to the crystal plasticity model and the equation of state.
The pullback magnitude is reproduced within 11\% of the experimental measurements and the time of the pullback minimum is reproduced within $\SI{4}{\nano\second}$ of experimental measurements.
Further, the first reload peak is reproduced within 9\% and $\SI{25}{\nano\second}$ of experimental measurements.
The onset stress, the start time, barrier height and the rate scale of the nucleation law are fitted to the $\SI{196}{\meter\per\second}$ experimental measurements and therefore agreement in $\SI{196}{\meter\per\second}$ simulated measurements with experiments is a calibrated result and not an independent prediction.
Further, modest deviations in the FSV from experimental measurements can be attributed to the lack of accuracy of the coalescence model and the bonded interface between the flyer and the target (beyond $\SI{2.8}{\micro\second}$).

The calibrated model is validated against the flyer-plate impact experiment at a flyer velocity of $\SI{214}{\meter\per\second}$.
All input parameters are identical to those used in the $\SI{196}{\meter\per\second}$ case, with the exception of the flyer velocity.
It is assumed that the same microstructural features lead to void nucleation in both cases so the same ML predictions are used in both cases.
The simulation gives a peak velocity 6\% lower than experimental measurements and a pullback magnitude 7\% lower than experimental measurements.
Further, the pullback minimum and the first reload arrive $\SI{35}{\nano\second}$ and $\SI{88}{\nano\second}$ later than experimental measurements.
The ringing after spall is captured well with some small deviations from the experiment in the same manner as in the $\SI{196}{\meter\per\second}$ case.
Further, the effect of the regularization length, $\ell$ and random seeds in the nucleation model is considered: for values of $\ell$ between $\SI{5.7}{\micro\meter}$ and $\SI{13.5}{\micro\meter}$ and four seeds at $\ell$ = $\SI{8}{\micro\meter}$, the pullback minimum and its time change by less than $\SI{0.3}{\meter\per\second}$ and $\SI{3}{\nano\second}$.
The first reload varies within 8\% of the experimental measurements.
Overall, for different $\ell$ values and random seeds, the FSV is accurately captured.

The simulations also resolve individual voids in the microstructure resolved within the spall plane.
Voids nucleate at the spall plane once the release fans interact with each other, and nearly all of them lie on grain boundaries, with several close to triple junctions.
Across four simulations that differ only in the random seed of the nucleation law, specific triple junctions and boundaries nucleate voids in every simulation, indicating that certain boundaries are more likely to nucleate voids than others.
Additionally, the intra granular void nucleates in the same grain in all simulations.
Only a small fraction of the boundaries nucleate voids although the entire spall region is under a similar tensile stress, indicating the dependence of void nucleation on grain boundary properties and microstructural features and not just local stresses.
The plastic work density is highest at the void rims during nucleation and then localizes in the ligaments between neighboring voids.
The voids grow and coalesce along these ligaments into a continuous damage region, which follows along grain boundaries over a part of its length and also cuts across grain interiors.
The damage model carries no information about the grain boundaries, so the path of the damage band is set by the nucleation sites and the plasticity in the regions around voids.
The tensile stress in the spall region relaxes to a small value once the damage band has formed.
Even though the four simulations have different damage patterns, they have similar FSVs, agreeing with experimental findings \cite{jones2018effect}.

Simulations with synthetic microstructures embedded in the spall region isolate the effect of the microstructure on the spall response.
In a single crystal, the predicted susceptibility field vanishes since there are no grain boundaries, indicating that the stress criterion alone contributes to void nucleation.
No voids nucleate during the simulation, and the FSV exhibits no spall signal.
The absence of a spall signal in the FSV is consistent with the higher spall strength of single tantalum crystals as compared to polycrystals.
However, in the single crystal case, voids can only nucleate via the stress criterion, since intragranular nucleation is not represented in susceptibility fields from the ML model.
Whether this represents single crystal spall accurately is not well established.

In the synthetic polycrystal with the same orientation distribution as the experimental material and a mean grain size of $\SI{5}{\micro\meter}$, the higher density of grain boundaries leads to a larger number of nucleation sites predicted by the ML model.
The radius of nucleated voids is reduced to $\SI{1.6}{\micro\meter}$ since grains are smaller as compared to experimental microstructures.
All other parameters are kept at the values of the experimental microstructure simulations.
The pullback signal is reproduced in these synthetic polycrystal simulations, but the spall region reaches complete separation rapidly and the amplitude of the late time ringing increases due to the nucleation of a large number of voids close to each other.
These results indicate that the pullback signal is  insensitive to the microstructure in the spall region but the late time ringing depends strongly on the number of nucleation sites and coalescence of nucleating voids.

We conclude by discussing the limitations of the work.
Because the simulations are 2D, out-of-plane effects (particularly effects of out-of-plane GB inclination) are not accounted for; one may reasonably expect to obtain a better match with full 3D simulations.
The transverse width of the domain is a very small part of the experimental target; although it appears to be fairly representative of the damage, the nucleation threshold is kept fairly low (leading to a high number of voids) so that the pullback signal is recovered, and the number of voids and the extent of coalescence are a consequence of this choice.
The flyer and target remain bonded throughout the simulation, which affects the FSV after the pullback minimum.
The microstructure embedded in the spall plane is pristine, so the simulated damage field is not compared against damaged regions of the recovered micrographs.
Intragranular void nucleation is not represented in the learned susceptibility field, so that the framework cannot predict spall accurately in single crystals or in grain interiors.
Nevertheless, the capacity of the model demonstrates the importance of microstructure and grain boundary awareness in simulations of this type.

\section*{Declarations}

\noindent\textbf{Funding}:
BR and ATA acknowledge support from the National Science Foundation CAREER program, grant number 2341922.

\noindent\textbf{Credit statements}:
ATA: Methodology, Validation, Formal analysis, Investigation, Writing - Original Draft. 
BR: Conceptualization, Writing - Review \& Editing, Supervision, Project administration, Funding acquisition

\noindent\textbf{Competing interests}:
The authors have no relevant financial or non-financial interests to disclose.

\noindent\textbf{Data availability}:
All data and code used for this paper are available upon request. 

\noindent\textbf{Declaration of generative AI and AI-assisted technologies in the manuscript preparation process}:
Statement: During the preparation of this work, the author used OpenAI Codex for copyediting and to provide constructive feedback on the manuscript; however, all text and content was human generated.
After using this tool/service, the author(s) reviewed and edited the content as needed and take(s) full responsibility for the content of the published article.

\printbibliography

@article{jones2018effect,
  title={Effect of peak stress and tensile strain-rate on spall in tantalum},
  author={Jones, David Robert and Fensin, Saryu Jindal and Martinez, Daniel Tito and Trujillo, Carl Patrick and Gray, GT},
  journal={Journal of Applied Physics},
  volume={124},
  number={8},
  year={2018},
  publisher={AIP Publishing},
  pages = {085901},
  doi = {10.1063/1.5045045},
}

@article{fensin2014some,
  title={Why are some Interfaces in Materials Stronger than others?},
  author={Fensin, Saryu J and Cerreta, Ellen K and III, GT Gray and Valone, Steven M},
  journal={Scientific reports},
  volume={4},
  number={1},
  pages={5461},
  year={2014},
  publisher={Nature Publishing Group UK London}
}

@article{gray2012high,
  title={High-strain-rate deformation: mechanical behavior and deformation substructures induced},
  author={Gray III, George T},
  journal={Annual Review of Materials Research},
  volume={42},
  number={1},
  pages={285--303},
  year={2012},
  publisher={Annual Reviews}
}

@article{curran1987dynamic,
  title={Dynamic failure of solids},
  author={Curran, Donald R and Seaman, L and Shockey, DA},
  journal={Physics reports},
  volume={147},
  number={5-6},
  pages={253--388},
  year={1987},
  publisher={Elsevier}
}

@article{kanel2010spall,
  title={Spall fracture: methodological aspects, mechanisms and governing factors},
  author={Kanel, GI},
  journal={International journal of fracture},
  volume={163},
  number={1},
  pages={173--191},
  year={2010},
  publisher={Springer}
}

@article{barker1972laser,
  title={Laser interferometer for measuring high velocities of any reflecting surface},
  author={Barker, LM and Hollenbach, RE},
  journal={Journal of Applied Physics},
  volume={43},
  number={11},
  pages={4669--4675},
  year={1972},
  publisher={American Institute of Physics}
}

@article{yang2022grain,
  title={Grain boundary effects on spall behavior of high purity copper cylinder under sweeping detonation},
  author={Yang, Yang and Huang, Jun-yi and Wang, Hai-min and Chen, Ji-xiong and Guo, Zhao-liang},
  journal={Journal of Central South University},
  volume={29},
  number={4},
  pages={1107--1117},
  year={2022},
  publisher={Springer}
}

@article{fensin2012influence,
  title={Influence of grain boundary properties on spall strength: Grain boundary energy and excess volume},
  author={Fensin, Saryu Jindal and Valone, Steven Michael and Cerreta, Ellen Kathleen and Gray, George Thompson},
  journal={Journal of Applied Physics},
  volume={112},
  number={8},
  year={2012},
  publisher={AIP Publishing}
}

@article{escobedo2011effects,
  title={Effects of grain size and boundary structure on the dynamic tensile response of copper},
  author={Escobedo, JP and Dennis-Koller, Darcie and Cerreta, EK and Patterson, BM and Bronkhorst, CA and Hansen, BL and Tonks, D and Lebensohn, RA},
  journal={Journal of Applied Physics},
  volume={110},
  number={3},
  year={2011},
  publisher={AIP Publishing}
}

@article{anantharanga2026predicting,
  title={Predicting grain boundary void nucleation sites in microstructure during shock loading with convolutional neural networks},
  author={Anantharanga, Abhijith Thoopul and Plummer, Jackson and Fensin, Saryu and Runnels, Brandon},
  journal={Acta Materialia},
  pages={122145},
  year={2026},
  publisher={Elsevier}
}

@article{tuler1968criterion,
  author  = {Tuler, F. R. and Butcher, B. M.},
  title   = {A criterion for the time dependence of dynamic fracture},
  journal = {International Journal of Fracture Mechanics},
  volume  = {4},
  number  = {4},
  pages   = {431--437},
  year    = {1968}
}

@article{johnson1981dynamic,
  author  = {Johnson, J. N.},
  title   = {Dynamic fracture and spallation in ductile solids},
  journal = {Journal of Applied Physics},
  volume  = {52},
  number  = {4},
  pages   = {2812--2825},
  year    = {1981}
}

@article{schmelzer2025statistical,
  author  = {Schmelzer, N. J. and Lieberman, E. J. and Chen, N. and Dunham, S. D. and Anghel, V. and Gray, G. T. and Bronkhorst, C. A.},
  title   = {Statistical evaluation of microscale stress conditions leading to void nucleation in the weak shock regime},
  journal = {International Journal of Plasticity},
  volume  = {188},
  pages   = {104318},
  year    = {2025},
  doi     = {10.1016/j.ijplas.2025.104318},
  publisher = {Elsevier},
}

@article{schmelzer2025thermo,
  author  = {Schmelzer, N. J. and Lieberman, E. J. and Gray, G. T. and Bronkhorst, C. A.},
  title   = {Thermodynamically consistent and microstructure informed porosity-based dynamic ductile damage model},
  journal = {Journal of the Mechanics and Physics of Solids},
  volume  = {205},
  pages   = {106336},
  year    = {2025}
}

@article{runnels2016analytical,
  title={An analytical model of interfacial energy based on a lattice-matching interatomic energy},
  author={Runnels, Brandon and Beyerlein, Irene J and Conti, Sergio and Ortiz, Michael},
  journal={Journal of the Mechanics and Physics of Solids},
  volume={89},
  pages={174--193},
  year={2016},
  publisher={Elsevier}
}

@article{zhang2023data,
  author  = {Zhang, Y. and Chen, N. and Bronkhorst, C. A. and Cho, H. and Argus, R.},
  title   = {Data-driven statistical reduced-order modeling and quantification of polycrystal mechanics leading to porosity-based ductile damage},
  journal = {Journal of the Mechanics and Physics of Solids},
  volume  = {179},
  pages   = {105386},
  year    = {2023}
}

@article{saha2026numerical,
  author  = {Saha, I. and Graham-Brady, L.},
  title   = {Numerical and data-driven modeling of spall failure in polycrystalline ductile materials},
  journal = {Computer Methods in Applied Mechanics and Engineering},
  volume  = {448},
  pages   = {118493},
  year    = {2026}
}

@article{versino2018computationally,
  author  = {Versino, D. and Bronkhorst, C. A.},
  title   = {A computationally efficient ductile damage model accounting for nucleation and micro-inertia at high triaxialities},
  journal = {Computer Methods in Applied Mechanics and Engineering},
  volume  = {333},
  pages   = {395--420},
  year    = {2018}
}

@article{brown2015microstructural,
  title={Microstructural effects on damage nucleation in shock-loaded polycrystalline copper},
  author={Brown, Andrew David and Wayne, Leda and Pham, Quan and Krishnan, Kapil and Peralta, Pedro and Luo, Sheng-Nian and Patterson, Brian M and Greenfield, Scott and Byler, Darrin and McClellan, Kenneth J and others},
  journal={Metallurgical and Materials Transactions A},
  volume={46},
  number={10},
  pages={4539--4547},
  year={2015},
  publisher={Springer}
}

@article{czarnota2006modelling,
  title={Modelling of nucleation and void growth in dynamic pressure loading, application to spall test on tantalum},
  author={Czarnota, Christophe and Mercier, S{\'e}bastien and Molinari, Alain},
  journal={International journal of fracture},
  volume={141},
  number={1},
  pages={177--194},
  year={2006},
  publisher={Springer}
}

@article{lee2026bayesian,
  title={Bayesian inference and uncertainty quantification for modeling of body-centered-cubic single crystals},
  author={Lee, Seunghyeon and Nguyen, Thao and Luscher, Darby J and Fensin, Saryu J and Carpenter, John S and Cho, Hansohl},
  journal={International Journal of Plasticity},
  pages={104678},
  year={2026},
  publisher={Elsevier}
}

@article{nguyen2017dislocation,
  title={A dislocation-based crystal plasticity framework for dynamic ductile failure of single crystals},
  author={Nguyen, Thao and Luscher, Darby Jon and Wilkerson, Justin W},
  journal={Journal of the Mechanics and Physics of Solids},
  volume={108},
  pages={1--29},
  year={2017},
  publisher={Elsevier},
  doi = {10.1016/j.jmps.2017.07.020},
}

@article{nguyen2021dynamic,
  title={Dynamic crystal plasticity modeling of single crystal tantalum and validation using Taylor cylinder impact tests},
  author={Nguyen, Thao and Fensin, Saryu J and Luscher, Darby J},
  journal={International Journal of Plasticity},
  volume={139},
  pages={102940},
  year={2021},
  publisher={Elsevier},
  doi = {10.1016/j.ijplas.2021.102940},
}

@article{remington2015microjoules,
  title={From microjoules to megajoules and kilobars to gigabars: Probing matter at extreme states of deformation},
  author={Remington, Bruce A and Rudd, Robert E and Wark, Justin S},
  journal={Physics of Plasmas},
  volume={22},
  number={9},
  year={2015},
  publisher={AIP Publishing}
}

@book{meyers1994dynamic,
  title={Dynamic behavior of materials},
  author={Meyers, Marc A},
  year={1994},
  publisher={John wiley \& sons},
  address = {New York},
}

@book{antoun2003spall,
  title={Spall fracture},
  author={Antoun, Tarabay and Curran, Donald R and Razorenov, Sergey V and Seaman, Lynn and Kanel, Gennady I and Utkin, Alexander V},
  year={2003},
  publisher={Springer}
}

@article{lebensohn2013modeling,
  title={Modeling void growth in polycrystalline materials},
  author={Lebensohn, Ricardo A and Escobedo, Juan P and Cerreta, Ellen K and Dennis-Koller, Darcie and Bronkhorst, Curt A and Bingert, John F},
  journal={Acta Materialia},
  volume={61},
  number={18},
  pages={6918--6932},
  year={2013},
  publisher={Elsevier},
  doi = {10.1016/j.actamat.2013.08.004},
}

@book{wilkins1999,
  author    = {Wilkins, M. L.},
  title     = {Computer Simulation of Dynamic Phenomena},
  publisher = {Springer},
  address   = {Berlin},
  year      = {1999},
  series = {Scientific Computation},
}

@article{benson1992,
  author    = {Benson, D. J.},
  title     = {Computational methods in {Lagrangian} and {Eulerian} hydrocodes},
  journal   = {Computer Methods in Applied Mechanics and Engineering},
  volume    = {99},
  number    = {2--3},
  pages     = {235--394},
  year      = {1992}
}

@book{davison2008,
  author    = {Davison, L.},
  title     = {Fundamentals of Shock Wave Propagation in Solids},
  publisher = {Springer},
  address   = {Berlin},
  year      = {2008}
}

@article{mie1903,
  author    = {Mie, G.},
  title     = {Zur kinetischen {Theorie} der einatomigen {K\"orper}},
  journal   = {Annalen der Physik},
  volume    = {316},
  number    = {8},
  pages     = {657--697},
  year      = {1903}
}

@book{clayton2019,
  author    = {Clayton, J. D.},
  title     = {Nonlinear Elastic and Inelastic Models for Shock Compression of Crystalline Solids},
  publisher = {Springer},
  address   = {Cham},
  year      = {2019},
  series = {Shock Wave and High Pressure Phenomena},
}

@book{marsh1980,
  editor    = {Marsh, S. P.},
  title     = {{LASL} Shock {Hugoniot} Data},
  series    = {Los Alamos Series on Dynamic Material Properties},
  publisher = {University of California Press},
  address   = {Berkeley, CA},
  year      = {1980}
}

@article{vonneumann1950,
  author    = {von Neumann, J. and Richtmyer, R. D.},
  title     = {A method for the numerical calculation of hydrodynamic shocks},
  journal   = {Journal of Applied Physics},
  volume    = {21},
  number    = {3},
  pages     = {232--237},
  year      = {1950}
}

@techreport{landshoff1955,
  author    = {Landshoff, R.},
  title     = {A numerical method for treating fluid flow in the presence of shocks},
  institution = {Los Alamos Scientific Laboratory},
  number    = {LA-1930},
  year      = {1955}
}

@article{wilkins1980,
  author    = {Wilkins, M. L.},
  title     = {Use of artificial viscosity in multidimensional fluid dynamic calculations},
  journal   = {Journal of Computational Physics},
  volume    = {36},
  number    = {3},
  pages     = {281--303},
  year      = {1980}
}

@article{caramana1998,
  author    = {Caramana, E. J. and Shashkov, M. J. and Whalen, P. P.},
  title     = {Formulations of artificial viscosity for multi-dimensional shock wave computations},
  journal   = {Journal of Computational Physics},
  volume    = {144},
  number    = {1},
  pages     = {70--97},
  year      = {1998}
}

@article{asaro1983,
  author  = {Asaro, R. J.},
  title   = {Micromechanics of crystals and polycrystals},
  journal = {Advances in Applied Mechanics},
  volume  = {23},
  pages   = {1--115},
  year    = {1983}
}

@article{austin2011,
  author  = {Austin, R. A. and McDowell, D. L.},
  title   = {A dislocation-based constitutive model for viscoplastic deformation of fcc metals at very high strain rates},
  journal = {International Journal of Plasticity},
  volume  = {27},
  number  = {1},
  pages   = {1--24},
  year    = {2011}
}

@book{kocks1975,
  author    = {Kocks, U. F. and Argon, A. S. and Ashby, M. F.},
  title     = {Thermodynamics and Kinetics of Slip},
  series    = {Progress in Materials Science},
  volume    = {19},
  publisher = {Pergamon Press},
  address   = {Oxford},
  year      = {1975}
}

@article{kothari1998,
  author  = {Kothari, M. and Anand, L.},
  title   = {Elasto-viscoplastic constitutive equations for polycrystalline metals: application to tantalum},
  journal = {Journal of the Mechanics and Physics of Solids},
  volume  = {46},
  number  = {1},
  pages   = {51--83},
  year    = {1998}
}

@article{lloyd2014shock,
  author  = {Lloyd, J. T. and Clayton, J. D. and Becker, R. and McDowell, D. L.},
  title   = {Simulation of shock wave propagation in single crystal and polycrystalline aluminum},
  journal = {International Journal of Plasticity},
  volume  = {60},
  pages   = {118--144},
  year    = {2014}
}

@article{luscher2017,
  author  = {Luscher, D. J. and Addessio, F. L. and Cawkwell, M. J. and Ramos, K. J.},
  title   = {A dislocation density-based continuum model of the anisotropic shock response of single crystal $\alpha$-cyclotrimethylene trinitramine},
  journal = {Journal of the Mechanics and Physics of Solids},
  volume  = {98},
  pages   = {63--86},
  year    = {2017}
}

@article{ma2004,
  author  = {Ma, A. and Roters, F.},
  title   = {A constitutive model for fcc single crystals based on dislocation densities and its application to uniaxial compression of aluminium single crystals},
  journal = {Acta Materialia},
  volume  = {52},
  number  = {12},
  pages   = {3603--3612},
  year    = {2004}
}

@article{orowan1940,
  author  = {Orowan, E.},
  title   = {Problems of plastic gliding},
  journal = {Proceedings of the Physical Society},
  volume  = {52},
  number  = {1},
  pages   = {8--22},
  year    = {1940}
}

@article{seppala2004effect,
  author  = {Sepp{\"a}l{\"a}, E. T. and Belak, J. and Rudd, R. E.},
  title   = {Effect of stress triaxiality on void growth in dynamic fracture of metals: A molecular dynamics study},
  journal = {Physical Review B},
  volume  = {69}, pages = {134101}, year = {2004},
  doi     = {10.1103/PhysRevB.69.134101}
}

@article{tang2012ductile,
  author  = {Tang, Y. and Bringa, E. M. and Meyers, M. A.},
  title   = {Ductile tensile failure in metals through initiation and growth of nanosized voids},
  journal = {Acta Materialia},
  volume  = {60}, number = {12}, pages = {4856--4865}, year = {2012},
  doi     = {10.1016/j.actamat.2012.05.030}
}

@article{molinari2005physical,
  author  = {Molinari, A. and Wright, T. W.},
  title   = {A physical model for nucleation and early growth of voids in ductile materials under dynamic loading},
  journal = {Journal of the Mechanics and Physics of Solids},
  volume  = {53}, number = {7}, pages = {1476--1504}, year = {2005}
}

@article{wilkerson2014dynamic,
  author  = {Wilkerson, J. W. and Ramesh, K. T.},
  title   = {A dynamic void growth model governed by dislocation kinetics},
  journal = {Journal of the Mechanics and Physics of Solids},
  volume  = {70}, pages = {262--280}, year = {2014}
}

@article{zhang2021phase,
  author  = {Zhang, H. and Peng, H. and Pei, X.-Y. and Li, P. and Tang, T.-G. and Cai, L.-C.},
  title   = {A phase-field model for spall fracture},
  journal = {Journal of Applied Physics},
  volume  = {129}, number = {12}, pages = {125903}, year = {2021},
  doi     = {10.1063/5.0043675}
}

@article{zhang2023phase,
  author  = {Zhang, H. and Peng, H. and Pei, X.-Y. and Wu, J.-Y. and Li, P. and Tang, T.-G. and Cai, L.-C. and Li, Y. and Liu, H.},
  title   = {Phase-field modeling of coupled spall and adiabatic shear banding and simulation of complex cracks in ductile metals},
  journal = {Journal of the Mechanics and Physics of Solids},
  volume  = {172}, pages = {105186}, year = {2023},
  doi     = {10.1016/j.jmps.2022.105186}
}

@article{han2024study,
  author  = {Han, H. and Wang, T. and Huang, G. and Liu, Z. and Zhuang, Z.},
  title   = {Study of the dynamic impact spalling of ductile materials based on {Gurson}-type phase-field model},
  journal = {International Journal of Plasticity},
  volume  = {181}, pages = {104106}, year = {2024},
  doi     = {10.1016/j.ijplas.2024.104106}
}

@article{runnels2025alamo, author={Runnels, B. and Agrawal, V. and Meier, M.}, title={The {Alamo} multiphysics solver for phase field simulations with strong-form mechanics and block structured adaptive mesh refinement}, journal={Journal of Open Source Software}, volume={10}, number={116}, pages={8581}, year={2025}, doi={10.21105/joss.08581}}

@article{zhang2019amrex, author={Zhang, W. and Almgren, A. and Beckner, V. and Bell, J. and Blaschke, J. and Chan, C. and Day, M. and Friesen, B. and Gott, K. and Graves, D. and others}, title={{AMReX}: a framework for block-structured adaptive mesh refinement}, journal={Journal of Open Source Software}, volume={4}, number={37}, pages={1370}, year={2019}}

@article{runnels2021massively, author={Runnels, B. and Agrawal, V. and Zhang, W. and Almgren, A.}, title={Massively parallel finite difference elasticity using block-structured adaptive mesh refinement with a geometric multigrid solver}, journal={Journal of Computational Physics}, volume={427}, pages={110065}, year={2021}}

@article{chen2020associating,
  author  = {Chen, Jie and Fensin, Saryu J.},
  title   = {Associating damage nucleation and distribution with grain boundary characteristics in {Ta}},
  journal = {Scripta Materialia},
  volume  = {187},
  pages   = {329--334},
  year    = {2020}
}

@article{nguyen2019role,
  author  = {Nguyen, Thao and Luscher, Darby J. and Wilkerson, Justin W.},
  title   = {The role of elastic and plastic anisotropy in intergranular spall failure},
  journal = {Acta Materialia},
  volume  = {168},
  pages   = {1--12},
  year    = {2019}
}

@article{remington2018spall,
  title={Spall strength dependence on grain size and strain rate in tantalum},
  author={Remington, TP and Hahn, EN and Zhao, S and Flanagan, R and Mertens, JCE and Sabbaghianrad, S and Langdon, Terence G and Wehrenberg, CE and Maddox, BR and Swift, DC and others},
  journal={Acta Materialia},
  volume={158},
  pages={313--329},
  year={2018},
  publisher={Elsevier}
}

@article{alessi2015, author={Alessi, R. and Marigo, J.-J. and Vidoli, S.}, title={Gradient damage models coupled with plasticity: variational formulation and main properties}, journal={Mechanics of Materials}, volume={80}, pages={351--367}, year={2015}}

@article{amor2009, author={Amor, H. and Marigo, J.-J. and Maurini, C.}, title={Regularized formulation of the variational brittle fracture with unilateral contact: numerical experiments}, journal={Journal of the Mechanics and Physics of Solids}, volume={57}, number={8}, pages={1209--1229}, year={2009}}

@article{borden2016, author={Borden, M. J. and Hughes, T. J. R. and Landis, C. M. and Anvari, A. and Lee, I. J.}, title={A phase-field formulation for fracture in ductile materials: finite deformation balance law derivation, plastic degradation, and stress triaxiality effects}, journal={Computer Methods in Applied Mechanics and Engineering}, volume={312}, pages={130--166}, year={2016}}

@article{bourdin2000, author={Bourdin, B. and Francfort, G. A. and Marigo, J.-J.}, title={Numerical experiments in revisited brittle fracture}, journal={Journal of the Mechanics and Physics of Solids}, volume={48}, number={4}, pages={797--826}, year={2000}}

@article{lemaitre1985, author={Lemaitre, J.}, title={A continuous damage mechanics model for ductile fracture}, journal={Journal of Engineering Materials and Technology}, volume={107}, number={1}, pages={83--89}, year={1985}}

@article{miehe2010, author={Miehe, C. and Hofacker, M. and Welschinger, F.}, title={A phase field model for rate-independent crack propagation: robust algorithmic implementation based on operator splits}, journal={Computer Methods in Applied Mechanics and Engineering}, volume={199}, number={45--48}, pages={2765--2778}, year={2010}}

@article{miehe2015, author={Miehe, C. and Hofacker, M. and Sch{\"a}nzel, L.-M. and Aldakheel, F.}, title={Phase field modeling of fracture in multi-physics problems. {Part II}. Coupled brittle-to-ductile failure criteria and crack propagation in thermo-elastic-plastic solids}, journal={Computer Methods in Applied Mechanics and Engineering}, volume={294}, pages={486--522}, year={2015}}

@article{miehe2016, author={Miehe, C. and Aldakheel, F. and Raina, A.}, title={Phase field modeling of ductile fracture at finite strains: a variational gradient-extended plasticity-damage theory}, journal={International Journal of Plasticity}, volume={84}, pages={1--32}, year={2016}}

@article{pham2011, author={Pham, K. and Amor, H. and Marigo, J.-J. and Maurini, C.}, title={Gradient damage models and their use to approximate brittle fracture}, journal={International Journal of Damage Mechanics}, volume={20}, number={4}, pages={618--652}, year={2011}}

@incollection{wilkins1964, author={Wilkins, M. L.}, title={Calculation of elastic-plastic flow}, booktitle={Methods in Computational Physics}, volume={3}, pages={211--263}, publisher={Academic Press}, address={New York}, year={1964}}

@article{fisher1948fracture,
  title={The fracture of liquids},
  author={Fisher, John C},
  journal={Journal of applied Physics},
  volume={19},
  number={11},
  pages={1062--1067},
  year={1948},
  publisher={American Institute of Physics}
}

@article{luo2012gas,
  author  = {Luo, S. N. and Jensen, B. J. and Hooks, D. E. and Fezzaa, K. and Ramos, K. J. and Yeager, J. D. and Kwiatkowski, K. and Shimada, T.},
  title   = {Gas gun shock experiments with single-pulse x-ray phase contrast imaging and diffraction at the {A}dvanced {P}hoton {S}ource},
  journal = {Review of Scientific Instruments},
  volume  = {83},
  pages   = {073903},
  year    = {2012},
  doi     = {10.1063/1.4733704}
}

@article{coakley2020femtosecond,
  author  = {Coakley, J. and Higginbotham, A. and McGonegle, D. and Ilavsky, J. and Swinburne, T. D. and Wark, J. S. and Rahman, K. M. and Vorontsov, V. A. and Dye, D. and Lane, T. J. and Boutet, S. and Koglin, J. and Robinson, J. and Milathianaki, D.},
  title   = {Femtosecond quantification of void evolution during rapid material failure},
  journal = {Science Advances},
  volume  = {6},
  number  = {51},
  pages   = {eabb4434},
  year    = {2020},
  doi     = {10.1126/sciadv.abb4434}
}

@article{miehe2015a,
  author  = {Miehe, C. and Sch{\"a}nzel, L.-M. and Ulmer, H.},
  title   = {Phase field modeling of fracture in multi-physics problems. {Part I}. {Balance} of crack surface and failure criteria for brittle crack propagation in thermo-elastic solids},
  journal = {Computer Methods in Applied Mechanics and Engineering},
  volume  = {294},
  pages   = {449--485},
  year    = {2015},
  doi     = {10.1016/j.cma.2014.11.016}
}

@article{ambati2015,
  author  = {Ambati, M. and Gerasimov, T. and De Lorenzis, L.},
  title   = {Phase-field modeling of ductile fracture},
  journal = {Computational Mechanics},
  volume  = {55},
  number  = {5},
  pages   = {1017--1040},
  year    = {2015},
  doi     = {10.1007/s00466-015-1151-4}
}

@article{gokuli2021multiphase,
  title={Multiphase field modeling of grain boundary migration mediated by emergent disconnections},
  author={Gokuli, Mahi and Runnels, Brandon},
  journal={Acta Materialia},
  volume={217},
  pages={117149},
  year={2021},
  publisher={Elsevier}
}

\end{document}